\documentstyle[twoside,epsfig]{article}

\catcode`\@=11
\long\def\@makefntext#1{
\protect\noindent \hbox to 3.2pt {\hskip-.9pt  
$^{{\eightrm\@thefnmark}}$\hfil}#1\hfill}		

\def\@makefnmark{\hbox to 0pt{$^{\@thefnmark}$\hss}}	
	
\def\ps@myheadings{\let\@mkboth\@gobbletwo
\def\@oddhead{\hbox{}
\rightmark\hfil\eightrm\thepage}   
\def\@oddfoot{}\def\@evenhead{\eightrm\thepage\hfil
\leftmark\hbox{}}\def\@evenfoot{}
\def\sectionmark##1{}\def\subsectionmark##1{}}

\oddsidemargin=\evensidemargin
\newcounter{sectionc}\newcounter{subsectionc}\newcounter{subsubsectionc}
\renewcommand{\section}[1] {\vspace{12pt}\addtocounter{sectionc}{1} 
\setcounter{subsectionc}{0}\setcounter{subsubsectionc}{0}\noindent 
	{\tenbf\thesectionc. #1}\par\vspace{5pt}}
\renewcommand{\subsection}[1] {\vspace{12pt}\addtocounter{subsectionc}{1} 
	\setcounter{subsubsectionc}{0}\noindent 
	{\bf\thesectionc.\thesubsectionc. {\kern1pt \bfit #1}}\par\vspace{5pt}}
\renewcommand{\subsubsection}[1] {\vspace{12pt}\addtocounter{subsubsectionc}{1}
	\noindent{\tenrm\thesectionc.\thesubsectionc.\thesubsubsectionc.
	{\kern1pt \tenit #1}}\par\vspace{5pt}}
\newcommand{\nonumsection}[1] {\vspace{12pt}\noindent{\tenbf #1}
	\par\vspace{5pt}}

\newcounter{appendixc}
\newcounter{subappendixc}[appendixc]
\newcounter{subsubappendixc}[subappendixc]
\renewcommand{\thesubappendixc}{\Alph{appendixc}.\arabic{subappendixc}}
\renewcommand{\thesubsubappendixc}
	{\Alph{appendixc}.\arabic{subappendixc}.\arabic{subsubappendixc}}

\renewcommand{\appendix}[1] {\vspace{12pt}
        \refstepcounter{appendixc}
        \setcounter{figure}{0}
        \setcounter{table}{0}
        \setcounter{lemma}{0}
        \setcounter{theorem}{0}
        \setcounter{corollary}{0}
        \setcounter{definition}{0}
        \setcounter{equation}{0}
        \renewcommand{\thefigure}{\Alph{appendixc}.\arabic{figure}}
        \renewcommand{\thetable}{\Alph{appendixc}.\arabic{table}}
        \renewcommand{\theappendixc}{\Alph{appendixc}}
        \renewcommand{\thelemma}{\Alph{appendixc}.\arabic{lemma}}
        \renewcommand{\thetheorem}{\Alph{appendixc}.\arabic{theorem}}
        \renewcommand{\thedefinition}{\Alph{appendixc}.\arabic{definition}}
        \renewcommand{\thecorollary}{\Alph{appendixc}.\arabic{corollary}}
        \renewcommand{\theequation}{\Alph{appendixc}.\arabic{equation}}
        \noindent{\tenbf Appendix \theappendixc #1}\par\vspace{5pt}}
\newcommand{\subappendix}[1] {\vspace{12pt}
        \refstepcounter{subappendixc}
        \noindent{\bf Appendix \thesubappendixc. {\kern1pt \bfit #1}}
	\par\vspace{5pt}}
\newcommand{\subsubappendix}[1] {\vspace{12pt}
        \refstepcounter{subsubappendixc}
        \noindent{\rm Appendix \thesubsubappendixc. {\kern1pt \tenit #1}}
	\par\vspace{5pt}}

\newcommand{\textlineskip}{\baselineskip=13pt}
\newcommand{\smalllineskip}{\baselineskip=10pt}

\def\eightcirc{
\begin{picture}(0,0)
\put(4.4,1.8){\circle{6.5}}
\end{picture}}
\def\eightcopyright{\eightcirc\kern2.7pt\hbox{\eightrm c}} 

\newcommand{\copyrightheading}[1]
	{\vspace*{-2.5cm}\smalllineskip{\flushleft
	{\footnotesize International Journal of Modern Physics E, #1}\\
	{\footnotesize $\eightcopyright$\, World Scientific Publishing
	 Company}\\
	 }}

\newcommand{\publisher}[2]{{\begin{center}\footnotesize\smalllineskip 
	Received #1\\
	Revised #2
	\end{center}
	}}

\def\abstracts#1#2#3{{
	\centering{\begin{minipage}{4.5in}\baselineskip=10pt\footnotesize
	\parindent=0pt #1\par 
	\parindent=15pt #2\par
	\parindent=15pt #3
	\end{minipage}}\par}} 

\renewenvironment{thebibliography}[1]
	{\frenchspacing
	 \ninerm\baselineskip=11pt
	 \begin{list}{\arabic{enumi}.}
        {\usecounter{enumi}\setlength{\parsep}{0pt}     
	 \setlength{\leftmargin 12.7pt}{\rightmargin 0pt} 
         \setlength{\itemsep}{0pt} \settowidth
	{\labelwidth}{#1.}\sloppy}}{\end{list}}

\newcounter{itemlistc}
\newcounter{romanlistc}
\newcounter{alphlistc}
\newcounter{arabiclistc}

\newcommand{\fcaption}[1]{
        \refstepcounter{figure}
        \setbox\@tempboxa = \hbox{\footnotesize Fig.~\thefigure. #1}
        \ifdim \wd\@tempboxa > 5in
           {\begin{center}
        \parbox{5in}{\footnotesize\smalllineskip Fig.~\thefigure. #1}
            \end{center}}
        \else
             {\begin{center}
             {\footnotesize Fig.~\thefigure. #1}
              \end{center}}
        \fi}

\newcommand{\tcaption}[1]{
        \refstepcounter{table}
        \setbox\@tempboxa = \hbox{\footnotesize Table~\thetable. #1}
        \ifdim \wd\@tempboxa > 5in
           {\begin{center}
        \parbox{5in}{\footnotesize\smalllineskip Table~\thetable. #1}
            \end{center}}
        \else
             {\begin{center}
             {\footnotesize Table~\thetable. #1}
              \end{center}}
        \fi}

\def\@citex[#1]#2{\if@filesw\immediate\write\@auxout
	{\string\citation{#2}}\fi
\def\@citea{}\@cite{\@for\@citeb:=#2\do
	{\@citea\def\@citea{,}\@ifundefined
	{b@\@citeb}{{\bf ?}\@warning
	{Citation `\@citeb' on page \thepage \space undefined}}
	{\csname b@\@citeb\endcsname}}}{#1}}

\newif\if@cghi
\def\cite{\@cghitrue\@ifnextchar [{\@tempswatrue
	\@citex}{\@tempswafalse\@citex[]}}
\def\citelow{\@cghifalse\@ifnextchar [{\@tempswatrue
	\@citex}{\@tempswafalse\@citex[]}}
\def\@cite#1#2{{$\null^{#1}$\if@tempswa\typeout
	{IJCGA warning: optional citation argument 
	ignored: `#2'} \fi}}
\newcommand{\citeup}{\cite}

\def\pmb#1{\setbox0=\hbox{#1}
	\kern-.025em\copy0\kern-\wd0
	\kern.05em\copy0\kern-\wd0
	\kern-.025em\raise.0433em\box0}

\def\fnt#1#2{\footnotetext{\kern-.3em
	{$^{\mbox{\scriptsize #1}}$}{#2}}}

\def\fpage#1{\begingroup
\voffset=.3in
\thispagestyle{empty}\begin{table}[b]\centerline{\footnotesize #1}
	\end{table}\endgroup}

\def\runninghead#1#2{\pagestyle{myheadings}
\markboth{{\protect\footnotesize\it{\quad #1}}\hfill}
{\hfill{\protect\footnotesize\it{#2\quad}}}}
\font\tenrm=cmr10
\font\tenit=cmti10 
\font\tenbf=cmbx10
\font\bfit=cmbxti10 at 10pt
\font\ninerm=cmr9
\font\nineit=cmti9
\font\ninebf=cmbx9
\font\eightrm=cmr8

\def\qed{\hbox{${\vcenter{\vbox{			
   \hrule height 0.4pt\hbox{\vrule width 0.4pt height 6pt
   \kern5pt\vrule width 0.4pt}\hrule height 0.4pt}}}$}}

\def\bsc{{\sc a\kern-6.4pt\sc a\kern-6.4pt\sc a}}	
\def\bflatex{\bf L\kern-.30em\raise.3ex\hbox{\bsc}\kern-.14em 
T\kern-.1667em\lower.7ex\hbox{E}\kern-.125em X} 

\begin{document}

\runninghead
{A.V.Korol, A.V.Solov'yov and W.Greiner}
{Total energy losses due to the radiation in an
acoustically based undulator$\ldots$} 

\normalsize\textlineskip
\thispagestyle{empty}
\setcounter{page}{1}

\copyrightheading{Vol. 9, No. 1 (2000) 77--105}

\vspace*{0.88truein}

\fpage{1}
\centerline{\bf TOTAL ENERGY LOSSES DUE TO THE RADIATION IN}   
\vspace*{0.035truein}
\centerline{\bf AN ACOUSTICALLY BASED UNDULATOR: THE UNDULATOR}
\vspace*{0.035truein}
\centerline{\bf AND THE CHANNELING RADIATION INCLUDED}         
\vspace*{0.37truein}
\centerline{\footnotesize ANDREI V. KOROL$^{\dag,\P,\|}$
\footnote{$^{\|}$E-mail: korol@rpro.ioffe.rssi.ru;
\  korol@th.physik.uni-frankfurt.de.},
ANDREY V. SOLOV'YOV$^{\ddag,\P,\S}$
\footnote{$^{\S}$E-mail: solovyov@rpro.ioffe.rssi.ru; 
\  solovyov@th.physik.uni-frankfurt.de.},
and WALTER GREINER$^{\P}$}
\vspace*{0.015truein}
\centerline{$^{\dag}$\footnotesize\it Department of Physics, St. Petersburg
State Maritime Technical University,}
\baselineskip=10pt
\centerline{\footnotesize\it Leninskii prospect 101, St. Petersburg 198262, Russia}
\baselineskip=10pt
\centerline{$^{\ddag}$\footnotesize\it A.F.Ioffe Physical-Technical
Institute of the Academy of Sciences of Russia,}
\baselineskip=10pt
\centerline{\footnotesize\it Polytechnicheskaya 26, St. Petersburg 194021, Russia}
\baselineskip=10pt
\centerline{$^{\P}$\footnotesize\it Institut f\"{u}r Theoretische
Physik der Johann Wolfgang Goethe-Universit\"{a}t,}
\baselineskip=10pt
\centerline{\footnotesize\it Robert-Mayer Str. 8-10, 60054 Frankfurt am Main, Germany}
\vspace*{0.225truein}
\publisher{(received date)}{(revised date)}

\vspace*{0.21truein}
\abstracts{
This paper is devoted to the investigation of the radiation energy
losses of an ultra-relativistic charged particle channeling along a
crystal plane which is periodically bent by a transverse acoustic
wave.  In such a system there are two essential mechanisms leading to
the photon emission.  The first one is the ordinary channeling
radiation.  This radiation is generated as a result of the transverse
oscillatory motion of the particle in the channel.  The second one is
the acoustically induced radiation.  This radiation is emitted because
of the periodic bending of the particle's trajectory created by the
acoustic wave.  The general formalism described in our work is
applicable for the calculation of the total radiative losses
accounting for the contributions of both radiation mechanisms. We
analyze the relative importance of the two mechanisms at various
amplitudes and lengths of the acoustic wave and the energy of the
projectile particle.  We establish the ranges of projectile particle
energies, in which total energy loss is small for the LiH, C, Si, Ge,
Fe and W crystals.  This result is important for the determination of
the projectile particle energy region, in which acoustically induced
radiation of the undulator type and also the stimulated photon
emission can be effectively generated. The latter effects have been
described in our previous works.}{}{}


\vspace*{1pt}\textlineskip	
\section{Introduction}\label{Introduction}

This paper is devoted to the investigation of the radiation energy
losses of an ultra-relativistic  charged particle
channeling along a crystal plane which is periodically bent by a
transverse acoustic wave. 
In such a system, there are two essential mechanisms leading
to the photon emission.
The first one is the ordinary channeling radiation.
This radiation is generated as a result of the transverse oscillatory
motion of the particle in the channel. This radiation mechanism 
was suggested in Refs. \cite{Kumakhov1,Barysh77} for linear crystals
and later studied in numerous theoretical and experimental works
(see e.g. Refs. \cite{Kumakhov2,Kniga,Baier}).
The second one is the acoustically induced radiation 
(AIR)\citeup{laser,lsr_rev}.
The AIR is emitted because of the periodic bending
of the particle's trajectory created by the acoustic wave. 
This mechanism is of particular interest, because
the AIR has all the features of the undulator radiation, including
the possibility of the stimulated photon emission\citeup{laser,lsr_rev}.

\begin{figure}[htbp]
\epsfig{file=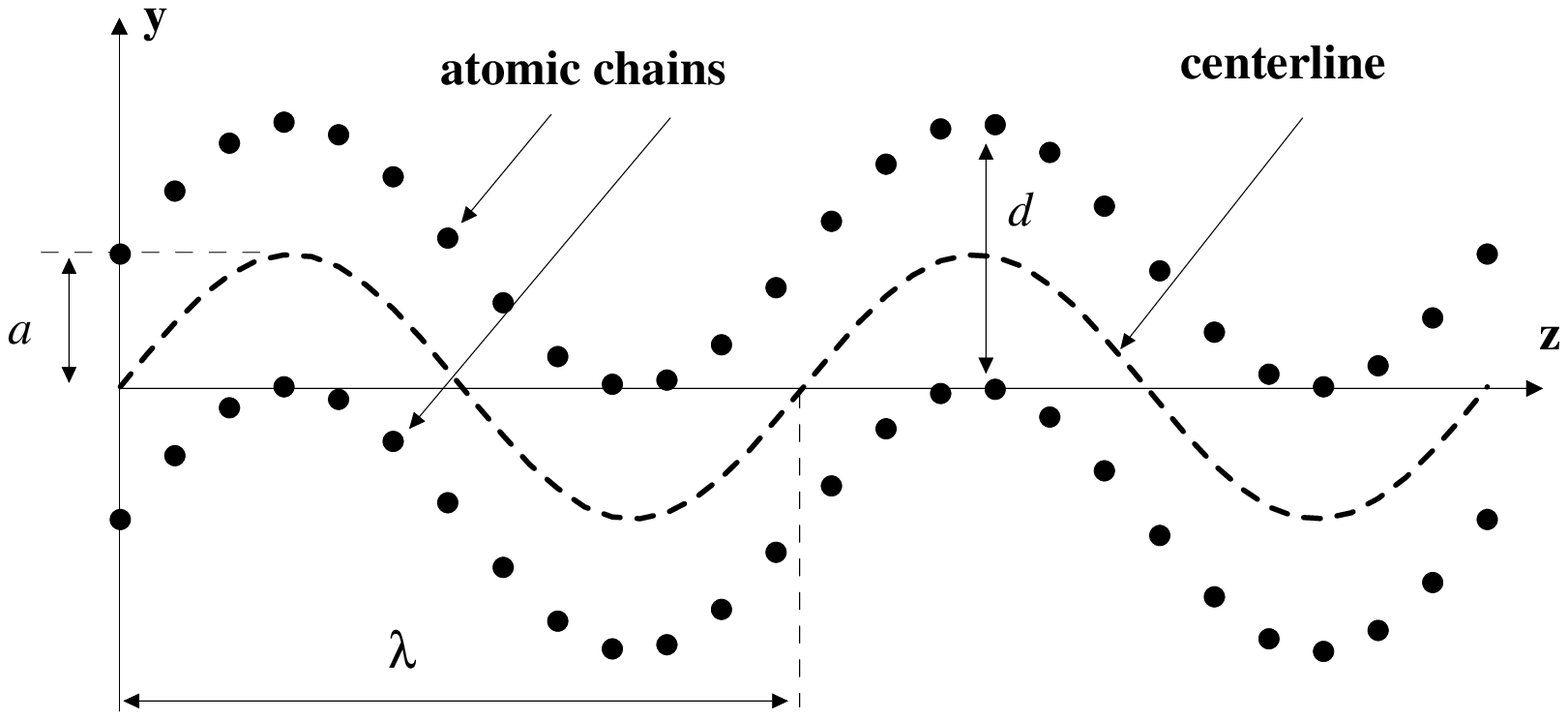,width=12cm,angle=0}
\vspace{0.5cm}
\fcaption{Schematic representation of the initially linear planar
channel bent by the transverse acoustic wave.
The notations are: $d$ is the channel width, $a, \lambda$ are the AW
amplitude and wavelength, respectively.}
\label{fig1}
\end{figure}

Since the mechanism of AIR generation was suggested only recently, let us
describe it here in more details.
This mechanism is illustrated in Fig.\ \ref{fig1}.

Under the action of a transverse acoustic wave propagating
along the $z$-direction, which defines the center line of an initially
straight channel (not plotted in the figure) the channel becomes
periodically bent.
Provided certain conditions are fulfilled\citeup{laser,lsr_rev},
the beam of positrons, which enters the
crystal at a small incident angle with respect to the curved
crystallographic plane, will penetrate through the crystal following
the bendings of its channel.
It results in the transverse oscillations of the beam particles
while travelling along the $z$ axis.
These oscillations become an effective source of spontaneous
radiation of undulator type due to the constructive interference
of the photons emitted from similar parts of the trajectory.
As demonstrated in Refs. \cite{laser,lsr_rev},
the number of oscillations can vary in a wide
range from a few up to a few thousands per $cm$ depending on the
the beam energy, the AW amplitude and wavelength 
the type of the crystal and the crystallographic plane.
In addition to the spontaneous photon emission by the undulator,
the scheme presented in Fig.\  \ref{fig1} leads to a possibility to
generate stimulated emission. This is due to the fact, that photons,
emitted at the
points of the maximum curvature of the trajectory, travel almost
parallel to the beam and, thus, stimulate the photon generation in
the vicinity of all successive maxima and minima of the trajectory.

As demonstrated in Refs. \cite{laser,lsr_rev}, the AIR can be well separated
from the ordinary channeling radiation, if certain conditions
on the amplitude and the length of the acoustic wave are fulfilled.
One of the criteria formulated in Refs. \cite{laser,lsr_rev} for 
the stable work of the AIR type of undulator
is the small energy loss of the beam of particles penetrating
through the crystal.

In this paper we describe general formalism for the calculation of
the total radiative energy loss accounting for the contributions
of both radiation  mechanisms.
We perform such a calculation for the first time.
We analyze the relative importance of the two mechanisms
at various amplitudes and lengths of the acoustic wave and the energy
of the projectile particle.
We establish the ranges of the projectile particle energy,
in which total radiative
energy loss is negligible for the LiH, C, Si, Ge, Fe and  W crystals.
This result is important for the determination of the
projectile particle energy region, in which acoustically induced radiation is
of the undulator type and also the stimulated photon emission can be
effectively generated.

An adequate approach to the problem of the radiation emission
by an ultra-relativistic particle moving in an external field
was developed by Baier and Katkov 
and was called by the authors the ``operator quasi-classical
method''. The details of that formalism can be found in
Ref. \cite{Baier}. We use this formalism to tackle our problem.

For convenience below we enlist the notations used throughout the paper. 

\begin{itemize}

\item
$\varepsilon$, $m$, and $\gamma=\varepsilon/m c^2$ are, respectively,
the energy, mass, and relativistic factor of a projectile,
$q$ is its charge measured in units of the
elementary charge $e$, $c$ is the velocity of light.

\item
$a_u$, $\lambda_u$ are  the AW amplitude and wave length. 
The undulator parameter equals to $p_u = \gamma\, \xi_u$,
where $\xi_u = 2\pi a_u/\lambda_u \ll 1$. 

\item
$\omega$ is the photon frequency, ${\bf n}$ is the unit vector in 
the direction of the emission.

\item
$L$ is the crystal thickness, $T \approx L/c$ is the time of flight
of the projectile through the crystal.

\item
$d$ is the interplanar spacing. It is assumed that $d$  satisfies the 
condition $d\ll \lambda_u$. 

\item
$U=U(\rho)$ is the interplanar potential, $\rho=[-d/2, +d/2]$ is the
distance from the midplane. The quantity $U_o$ stands for the maximum
value of $U(\rho)$.

\item
$C$ stands for the factor 
$\varepsilon/(R_{min}\, q e U^{\prime}_{max})\approx
\varepsilon\,d/(R_{min}\, 2 q e U_o)$, 
where $R_{min} = (k_u^2 a_u)^{-1}$
is the minimum curvature radius of an acoustically bent channel,
and $k_u = 2\pi/\lambda_u$, 
$U^{\prime}_{max} \approx  2 U_o/d$ is the maximum gradient of the
interplanar field. 

\item
$a_c$, $\lambda_c$ are, respectively, the amplitude and wave  
length characterizing the channeling motion. 
The corresponding undulator parameter reads as $p_c = \gamma\, \xi_c$. 
The $\xi$-parameter is defined as  $\xi_c = 2\pi a_c/\lambda_c$. 
More details on these parameters are given in the text.

\item
$\alpha= e^2/ \hbar\, c \approx 1/137$ is the fine structure constant,
$r_e = e^2/(m_e c^2) = 2.818 \times 10^{-13}$ cm is the electron classical 
radius.

\end{itemize}

\section{Quasi-classical formalism  for the  radiative energy loss}
\label{Radiation}

The energy losses, $\Delta\, E$, 
due to the emission of photons by a charged projectile moving in an
external field are defined as
\begin{equation}
\Delta\, E = \int_0^{\varepsilon/\hbar}{\rm d} \omega 
\int {\rm d} \Omega_{\bf n}
 { {\rm d} E_{\omega}({\bf n}) \over {\rm d} \omega {\rm d}
\Omega_{\bf n}}\ .
\label{1.1}
\end{equation}

Within the framework of the quasi-classical approach\citeup{Baier} the
distribution of the energy radiated by an ultra-relativistic particle 
(of a spin $s=1/2$) in given direction ${\bf n}$ 
and summed over the polarizations of the photon and the projectile,
is given by the following expression, which is written up to 
the terms $\gamma^{-2}$:
\begin{equation}
{ {\rm d} E_{\omega}({\bf n}) \over {\rm d} \omega {\rm d} \Omega_{\bf n}} =
\hbar\,\alpha \, { q^2\,\omega^2  \over 4 \pi^2 } \, 
\int_0^{T}\, {\rm d} t_1\ \int_0^{T}\, {\rm d} t_2 \
{\rm e}^{{\rm i} \omega^{\prime} \varphi(t_1,t_2)}
\ f(t_1,t_2).
\label{1.2}
\end{equation}
The functions $ \varphi(t_1,t_2)$ and $f(t_1,t_2)$ equal to
\begin{eqnarray}
 \varphi(t_1,t_2)& = &t_1 - t_2 - {1 \over c}\,
{\bf n}\cdot ({\bf r}_1 - {\bf r}_2)\ ,
\label{1.3} \\
 f(t_1, t_2)&=& {1 \over 2}
\biggl\{
\left( 1+(1+u)^2 \right)\,
\left(
{ {\bf v}_1\cdot {\bf v}_2 \over c^2} -1
\right)
+{u^2 \over \gamma^2}
\biggr\}.
\label{1.4}
\end{eqnarray}
The notations used are 
${\bf r}_{1,2}= {\bf r}(t_{1,2})$, ${\bf v}_{1,2}= {\bf v}(t_{1,2})$, 
with ${\bf r}$ and ${\bf v}$ standing for projectile's radius vector 
and  velocity, respectively.

Expression (\ref{1.4}) looks almost like the classical
formula\citeup{Baier},  although with quantum corrections:
\begin{equation}
\omega\longrightarrow
\omega^{\prime} =
{ \varepsilon \over
\varepsilon - \hbar \omega}\, \omega,
\qquad
u=
{ \hbar \omega \over
\varepsilon - \hbar \omega}\ ,
\label{1.5}
\end{equation}
which take into account the radiative recoil.

Expressing the photon frequency and the quantity $\omega^{\prime}$ 
via the dimensionless variable $u$,
$\hbar \omega = \varepsilon\, u/(1+u)$, 
$\hbar \omega^{\prime} = u\, \varepsilon$,  
and taking into account the relation
$\omega^2 {\rm d} \omega = \left({\varepsilon/\hbar}\right)^3\, 
u^2/(1+u)^4\, {\rm d} u$,
one obtains the following general expression for the relative energy
losses:
\begin{eqnarray}
{ \Delta\, E \over \varepsilon} 
&=& 
\alpha \, { q^2 \over 4 \pi^2 } \,
\left({\varepsilon \over \hbar}\right)^2\ 
\int_0^{\infty} {u^2 {\rm d} u \over (1+u)^4}
\nonumber\\
&\times&
\int {\rm d} \Omega_{\bf n}
\int_0^{T}\int_0^{T} {\rm d} t_1  {\rm d} t_2 \,
\exp\left({\rm i}\, {\varepsilon u \over \hbar}\, \varphi(t_1,t_2)\right)
\,f(t_1,t_2) .
\label{1.8}
\end{eqnarray}

Let us transform the functions $f(t_1,t_2)$ and $\varphi(t_1,t_2)$ 
retaining the terms of orders up to $\gamma^{-2}$ and omitting the 
higher-order terms.

To evaluating the factor ${\bf v}_1\cdot {\bf v}_2/ c^2 -1$ 
from (\ref{1.4}) one makes use of Eqs. (\ref{app.A14}) and
(\ref{app.A15}). 
Then, neglecting the term $\Delta \dot{z}(t_1)\,\Delta \dot{z}(t_2)$,
one gets 
\begin{equation}
{ {\bf v}_1\cdot {\bf v}_2 \over c^2} -1
\approx
- {1 \over \gamma^2} 
- 
{1 \over 2}\, 
\left({ v_y(t_1) \over c} - {v_y(t_2) \over c} \right)^2.
\label{1.9}
\end{equation}
Substituting this expression into (\ref{1.4}) we obtain
\begin{equation}
 f(t_1, t_2)= - {1 \over 2}
\biggl\{
{ 1+(1+u)^2 \over 2}\,
\left({ v_y(t_1) \over c} - {v_y(t_2) \over c} \right)^2
+
{2 (1+u) \over \gamma^2}
\biggr\}.
\label{1.10}
\end{equation}

To transform the phase function $\varphi(t_1,t_2)$ from (\ref{1.3})
let us first write the trajectory ${\bf r}(t)$ in the following form:
\begin{equation}
{\bf r}(t) =
{\bf e}_z\cdot\left(ct + \Delta z(t)\right)
+
{\bf e}_y\cdot y(t).
\label{1.11}
\end{equation}
For an arbitrary interplanar potential the function $\Delta z(t)$
can be presented in the form (see \ref{motion})
\begin{equation}
\Delta z(t) =
- c\, t\, 
\left[ 
{1 \over 2 \gamma^2} + {\xi_u^2 \over 4} + {\xi_c^2 \over 4}
\right]
+
\Delta_z(t)\ ,
\label{1.12}
\end{equation}
where the parameters $\xi_u^2$ and $\xi_c^2$ are related to the
mean-square values of the transverse velocities of, respectively, the
undulator and the channeling motions 
\begin{equation}
{\xi_u^2 \over 2} = 
\overline{\left({1 \over c} {{\rm d} y_u \over {\rm d} t}\right)^2},
\qquad 
{\xi_c^2 \over 2} = 
\overline{\left({1 \over c} {{\rm d} y_c \over {\rm d} t}\right)^2}.
\label{1.13b}
\end{equation}

For the undulator motion (i.e. the motion along the centerline of the
acoustically bent crystal, $y_u(t) = a_u\, \sin(2\pi c t/\lambda_u)$) 
the first relation from (\ref{1.13b}) produces the result 
$\xi_u = 2\pi a_u/\lambda_u$. 
Analogously, the parameter $\xi_c$ may be written as
$\xi_c = 2\pi a_c/\lambda_c$, where $a_c$ is the amplitude (mean) of the
channeling oscillations $a_c \leq d/2$ and $\lambda_c$ is the period
(mean) of the channeling oscillatory motion.
To estimate the magnitude of $\xi_c$ one can do the following:
$\xi_c \sim a_c/(c\, \tau_c)$ with $\tau_c \sim \sqrt{m \gamma d^2/ q e
U_o}$ standing for the period of the channeling oscillations. 
Hence $\xi_c^2 \sim q e U_o/ \varepsilon  \ll 1$.

The term $\Delta_z(t)$ in the order of magnitude equals to
$\Delta_z(t) = O(\xi_u^2,\ \xi_c^2,\ \xi_u \xi_c)$.
It contains only oscillatory terms which satisfy the condition
$\overline{\Delta_z(t)} =0$
if the averaging is carried out over the interval 
$\Delta T > \lambda_u/c,\ \tau_c$.

Now, to write down the term ${\bf n}\cdot ({\bf r}_1 - {\bf r}_2)$
from (\ref{1.3}) let us notice that for an ultra-relativistic particle
the radiation occurs into a narrow cone with the axis along the
$z$-direction. 
The width of the cone is defined by three parameters, 
$\gamma^{-2}$, $\xi_c^2$, and $\xi_u^2$ and is equal to
\begin{equation}
\theta_{max} \sim {\rm max}(\gamma^{-2}, \ \xi_u^2,\  \xi_c^2) \ll 1.
\label{1.17}
\end{equation}

The relations established above allow to write down the following 
expression for $\varphi(t_1,t_2)$ which explicitly accounts for all 
the terms of orders $\gamma^{-2}$, $\xi_c^2$, and $\xi_u^2$:
\begin{eqnarray}
\varphi(t_1,t_2) 
& =& 
\varphi_o(t_1,t_2) + \Delta \varphi(t_1,t_2)\ ,
\label{1.18a} \\
\varphi_o(t_1,t_2)
& =& 
 \kappa^2 \tau -
{1 \over c} \, (\Delta_z(t_1) - \Delta_z(t_2))\  ,
\label{1.18b}\\
\Delta \varphi(t_1,t_2)
& =& 
 {\tau \over 2}\, \theta^2 - 
\theta\, {y(t_1) - y(t_2) \over c} \, \cos\phi\ ,
\label{1.18c}
\end{eqnarray}
where the following notations are introduced:
\begin{equation}
\kappa^2 = 
{1 \over 2 \gamma^2} + {\xi_u^2 \over 4} + {\xi_c^2 \over 4},
\qquad
\tau = t_1 - t_2.
\label{1.19}
\end{equation}
When writing (\ref{1.18c}) we took into account that
$y(t_1) - y(t_2) \sim O(\xi_u, \,\xi_c)\, c\, \tau$ which results in 
$\sin\theta\, (y(t_1) - y(t_2)) =\theta\, (y(t_1) - y(t_2))$.

Using Eqs. (\ref{1.10}) and (\ref{1.18c}) in (\ref{1.8})
one gets
\begin{eqnarray}
{ \Delta\, E \over \varepsilon} &=& 
-{ \alpha \, q^2 \over 8 \pi^2 } \,
\left({\varepsilon \over \hbar}\right)^2\ 
\int_0^{T}\, {\rm d} t_1\ \int_0^{T}\, {\rm d} t_2 \
\int_0^{\infty} {u^2 {\rm d} u \over (1+u)^4}
\, \exp\left({\rm i} {\varepsilon u \over \hbar}\,\varphi_o(t_1,t_2)\right)
\nonumber
\\
\label{1.21}
&&\times
\biggl\{
{ 1+(1+u)^2 \over 2}\,
\left({ v_y(t_1) \over c} - {v_y(t_2) \over c} \right)^2
+
{2 (1+u) \over \gamma^2}
\biggr\}
\\
&&\times
\int {\rm d} \Omega_{\bf n}
\exp\left({\rm i}\, {\varepsilon u \over \hbar c}\, 
\Delta \varphi(t_1,t_2)\right).
\nonumber
\end{eqnarray}

Due to the relation (\ref{1.17}) the main contribution to 
the integral over $\Omega_{\bf n}=(\theta, \phi)$ comes from the
region $\theta \ll 1$. 
Therefore, one may write
\begin{equation}
I \equiv
\int {\rm d} \Omega_{\bf n}
\exp\left({\rm i}\, {\varepsilon u \over \hbar c}\, 
\Delta \varphi(t_1,t_2)\right) 
=
\int_0^{\infty} \theta {\rm d} \theta \
{\rm e}^{{\rm i} \omega^{\prime} \tau \theta^2 /2}
\int_0^{2\pi} {\rm d} \phi \
{\rm e}^{- {\rm i} \omega^{\prime} \Delta y \theta \cos\phi/c}.
\label{1.22}
\end{equation}
For short the notations $\omega^{\prime} =u \varepsilon/\hbar$
(see (\ref{1.5})) and $\Delta y = y(t_1) - y(t_2)$ were used.

The integrals over $\phi$ and $\theta$ are carried out by using the 
formulae\citeup{Abramowitz,Ryzhik} 
\begin{equation}
\int_0^{2\pi} {\rm d} \phi \
{\rm e}^{{\rm i} z \cos\phi}
= 2\pi\, J_0(z),
\quad
\int_0^{\infty} \exp\left({\rm i} a x^2\right)\, J_0(bx)\, x\, {\rm d} x 
= {{\rm i} \over 2 a}\,\exp\left(-{\rm i} {b^2\over 4a} \right),
\label{1.23}
\end{equation}
where $J_0(z)$ is the Bessel function of order 0.

Applying these integrals to (\ref{1.22}) one gets
\begin{equation}
I =
{\hbar \over \varepsilon u}\, 
{2\pi\, {\rm i} \over \tau}\, 
\exp\left[ -{\rm i} {\varepsilon u \over \hbar}\,\tau
\left( 1 + {(y(t_1)-y(t_2))^2 \over 2c^2\tau^2}
\right)\right].
\label{1.25}
\end{equation}

When substituting this result into (\ref{1.21}) we first introduce the
quantity
\begin{equation}
\zeta =
{\varepsilon \over \hbar}\, \tau\, 
\left\{
\kappa^2 -
{\Delta_z(t_1) - \Delta_z(t_2) \over c\, \tau} 
-
{1 \over 2}\, 
\left({ y(t_1)-y(t_2) \over c \tau} \right)^2
\right\}.
\label{1.26}
\end{equation}

Hence
\begin{eqnarray}
{ \Delta\, E \over \varepsilon} 
&=& 
-{\rm i} { \alpha \, q^2 \over 4 \pi} \,
\left({\varepsilon \over \hbar}\right)^2\ 
\int_0^{T} \int_0^{T} \, {{\rm d} t_1\, {\rm d} t_2 \over \tau}
\int_0^{\infty} {u^2 {\rm d} u \over (1+u)^4}\,  
\exp\left({\rm i} \zeta u\right)
\nonumber\\
&&\times
\biggl\{
{ 1+(1+u)^2 \over 2}\,
\left({ v_y(t_1) \over c} - {v_y(t_2) \over c} \right)^2
+
{2 (1+u) \over \gamma^2}
\biggr\}.
\label{1.27}
\end{eqnarray}

Let us demonstrate that the principal contribution to (\ref{1.27}) 
comes from the region $|\zeta|<1$.
To do this we first evaluate the integral over $u$ and then analyze
the result.
Making a substitution 
$\exp{{\rm i} \zeta u} \longrightarrow {\rm i}\, \sin \zeta u$ 
in the integrand (which does not affect the three-fold integral on the
right-hand side) 
one is left with two basic integrals
\begin{eqnarray}
j_1 &=& { 1 \over 2}\,
\int_0^{\infty}\, {u^2\, {\rm d} u \over (1+u)^4}
\, \left(1+(1+u)^2\right)\,
\sin \zeta u\ ,
\label{1.28} \\
j_2 &=& 2\,
\int_0^{\infty}\, {u\, {\rm d} u \over (1+u)^3}\,
\sin \zeta u\ ,
\label{1.29}
\end{eqnarray}
the evaluation of which is elementary but lengthly. The
result is 
\begin{eqnarray}
j_1 &=& { 1 \over 2}\,
\left[ {\zeta \over 3} + 
\left(1 - {\zeta^2 \over 2}\right)\, f(\zeta)
-
\left(\zeta - {\zeta^3 \over 6}\right)\, g(\zeta)
\right]\ ,
\label{1.30a} \\
j_2 &=& - \zeta +  
\zeta^2\, f(\zeta) + 2 \zeta\,g(\zeta)\ , 
\label{1.30b}
\end{eqnarray}
where the functions $f(\zeta)$ and $g(\zeta)$ stand for the integrals:
\begin{equation}
f(\zeta) = 
\int_0^{\infty}\, {\sin \zeta u  \over 1+u}\, {\rm d} u,
\qquad
g(\zeta) = 
\int_0^{\infty}\, {\cos \zeta u \over 1+u}\, {\rm d} u 
\label{1.32b}.
\end{equation}
 
For $|\zeta|>1$ the expansions for $f(\zeta)$ and $g(\zeta)$ one finds in
Ref. \cite{Abramowitz}. Using them one gets:
\begin{equation}
j_1 \approx {6 \over \zeta^3},
\qquad
j_2 \approx {12 \over \zeta^3}.
\label{1.33}
\end{equation}
Large values of $\zeta$ correspond to the $\tau$-values 
$|\tau| \gg \hbar/\varepsilon\, \kappa^2$ (see (\ref{1.26})). 
With the expressions (\ref{1.33}) taken into account 
the integrand in (\ref{1.27}) behaves as $\sim \tau^{-4}$.
Thus, the range of large $|\zeta|$ (or $|\tau|$) does not contribute
effectively to the integral (\ref{1.27}). 

The leading contribution to the energy losses (\ref{1.27}) comes from
the range $|\zeta| < 1$ which corresponds to
\begin{equation}
|t_1-t_2| \equiv |\tau| < {\varepsilon\, \kappa^2 \over \hbar}.
\label{1.34}
\end{equation}

\section{Energy losses in the case of harmonic interplanar potential}
\label{harm}

Instead of evaluating the remaining integrals in (\ref{1.27}) in the
general case we will consider the harmonic approximation for the
interplanar potential. It allows to carry out all the calculations
explicitly.  The final result can be generalized to the case of an
arbitrary interplanar potential.

\subsection{Evaluation of the formula for the energy losses.}

The harmonic potential, expressed in terms of the relative distance 
from the midplane, $\tilde{y}$, can be written in the form
\begin{equation}
U(\tilde{y}) = 4\, U_o\,{\tilde{y}^2 \over d^2}.
\label{2.0}
\end{equation}
The coefficient is chosen to satisfy $U(\pm d/2) = U_o$, with $U_o$
standing for the maximum value of the potential.

Assuming the following strong inequality 
(see \ref{harmonic})
\begin{equation}
{\Omega_c \over \Omega_u} \gg 1,
\label{2.00}
\end{equation}
which is valid almost in all cases,
the dependences $\Delta_z(t)$ and $y(t)$ read as (compare with 
(\ref{app.A17}) and (\ref{app.A20}))  
\begin{eqnarray}
y(t)&=& a_u\, \sin \Omega_u t + a_c\, \sin (\Omega_c t + \phi_0),
\label{2.1} \\
\Delta_z(t) &=& -{1 \over 16 \pi} \, 
\biggl\{
\lambda_u\,\xi_u^2\, \sin 2\Omega_u t +
\lambda_c\, \xi_c^2\, \sin (2\Omega_c t + 2\phi_0 ),
\label{2.2} \\
&&+ 4\,\lambda_c\, \xi_u\, \xi_c\,
\left[ 
\cos \left((\Omega_c + \Omega_u)t + \phi_0\right)
+
\cos \left((\Omega_c - \Omega_u)t + \phi_0\right)
\right]
\biggr\}.
\nonumber 
\end{eqnarray}
In these formulas the subscripts ``u'' and ``c'' indicate that a quantity
is related to the undulator motion (the index ``u'') or to the
channeling motion (the index ``c'').

The parameters  $a_u$, $\lambda_u$, $\xi_u=2\pi \, a_u/\lambda_u$, and
$\Omega_u= 2\pi \, c/\lambda_u$ are explained in \ref{Introduction}.
 
The quantities characterizing the channeling motion, which are:
the frequency of the channeling
oscillations $\Omega_c$, the wave length of one oscillation
$\lambda_c$, and the parameter $\xi_c=2\pi\, a_c/\lambda_c$, 
are conveniently expressed through the parameter $\mu$ (see Eq. 
(\ref{app.A18a}):
\begin{equation}
\Omega_c = {2 \, \mu \, c \over d},
\qquad
\lambda_c = {\pi \,d  \over  \mu},
\qquad
\xi_c = {2 \, \mu \, a_c \over  d}.
\label{2.4}
\end{equation}

It is also worth noting that expressions (\ref{2.1}) and (\ref{2.2})
embrace both the $a_u \gg d$ and $a_u \ll d$ regions. 
In the latter case $\xi_u=0$, and the formulae produce the result for the
channeling motion in a linear channel. 

To evaluate the integral (\ref{1.27}) let us first analyze the
functions 
$\left(\Delta_z(t_1) - \Delta_z(t_2)\right)/c\, \tau$,
$\left(y(t_1)-y(t_2)\right)^2/ (c \tau)^2 $ (see (\ref{1.26})) 
and $\left(v_y(t_1)  - v_y(t_2)\right)^2/ c^2$ from
the integrand.

It can be easily verified that when substituted into 
$\left(\Delta_z(t_1) - \Delta_z(t_2)\right)/c\, \tau$ 
the first term from the right-hand side of (\ref{2.2}) results in the
term of the order $\xi_u^2$, the second one produces the term 
$\sim\xi_c^2$ and the last one gives the $\sim\xi_c\,\xi_u$ term.
To illustrate this let us consider the contribution of the first term
from (\ref{2.2}):
\begin{equation}
{\lambda_u\,\xi_u^2 \over 16 c\,\pi} \, 
\left| {\sin 2\Omega_u t_1 - \sin 2\Omega_u t_2 \over t_1 - t_2}\right|
=
{\lambda_u\,\xi_u^2 \, \Omega_u \over 8 c\,\pi} \, 
\left| {\sin \Omega_u \tau \over \Omega_u\, \tau}\, 
 \cos\Omega_u (t_1 +t_2) \right| \leq 
{\xi_u^2 \over 4}.
\label{2.5}
\end{equation}

Next consideration regarding the difference 
$\left(\Delta_z(t_1) - \Delta_z(t_2)\right)/c\, \tau$ 
is that for any fixed value of $\tau$ this function is highly
oscillatory because of the factors of the type
$\cos\Omega_u (t_1 +t_2)$, $\cos\Omega_c (t_1 +t_2)$,
and/or $\sin\left( (\Omega_c \pm \Omega_u) (t_1 +t_2)/2 + \phi_0\right)$.
It leads to the relation
$\overline{ \Delta_z(t_1) - \Delta_z(t_2)}=0 $
for any value of $\tau$.
The leading value of the integral (\ref{1.27}) will not be changed
if in (\ref{1.26}) one omits these highly oscillatory terms.

By using similar arguments let substitute the function
$\left(y(t_1)-y(t_2)\right)^2/ (c \tau)^2 $ from (\ref{1.26}) with its
non-oscillatory part. 
Making use of (\ref{2.1}) one gets
\begin{equation}
{1 \over 2}\, \left({y(t_1)-y(t_2) \over c\,\tau} \right)^2
\longrightarrow
{1 \over 2}\,
\overline{\left({y(t_1)-y(t_2) \over c\,\tau} \right)^2}
=
{\xi_u^2 \over 4}\,\left( {\sin \eta_u \over \eta_u} \right)^2
+
{\xi_c^2 \over 4}\,\left( {\sin \eta_c \over \eta_c} \right)^2,
\label{2.7}
\end{equation}
where $\eta_{u,c}=\Omega_{u,c}\, \tau/2$.

The main contribution to the integral (\ref{1.27}) comes from the
region of small $\tau$ (see (\ref{1.34})), therefore, one may use
$(\sin \eta_{u,c}/\eta_{u,c})^2 = 1 - \eta_{u,c}^2/3$.

Thus, the function $\zeta$ defined in (\ref{1.26}) can be substituted
with
\begin{equation}
\zeta 
\longrightarrow
x\,\tau + a\,\tau^3, 
\label{2.8}
\end{equation}
where the following short-hand notations are introduced
\begin{equation}
x= {\varepsilon \over 2 \gamma^2 \hbar},
\qquad
a= {\varepsilon \over 2 \hbar}\,
{\xi_u^2 \Omega_{u}^2 + \xi_c^2 \Omega_{c}^2 \over 24}.
\label{2.9}
\end{equation}

The last function to be transformed is 
$\left(v_y(t_1)  - v_y(t_2)\right)^2/ c^2$ from  the integrand in 
(\ref{1.27}). 
The non-oscillatory part of it reads as
\begin{equation}
 \left({v_y(t_1)  - v_y(t_2) \over c} \right)^2
\longrightarrow
{\xi_u^2 \Omega_{u}^2 + \xi_c^2 \Omega_{c}^2 \over 2}\, \tau^2.
\label{2.10}
\end{equation}

Substituting (\ref{2.8})--(\ref{2.10}) into
(\ref{1.27}), and introducing the integration variable $\tau$ one
obtains
\begin{eqnarray}
{ \Delta\, E \over \varepsilon} &=& 
-{\rm i}\, { \alpha \, q^2 \over 4 \pi} \,
{\varepsilon \over \hbar}\,
\int_0^{T} {\rm d} t 
\int_0^{\infty} {u^2 {\rm d} u \over (1+u)^4}\,  
\int_{-t}^{t} \, {{\rm d} \tau \over \tau}\, 
\exp\left[{\rm i}\, u (x\tau+a\tau^3)\right]
\nonumber\\
&&\times
\biggl\{
{ 1+(1+u)^2 \over 2}\,
{\xi_u^2 \Omega_{u}^2 + \xi_c^2 \Omega_{c}^2 \over 4} 
+
{2(1+u) \over \gamma^2}
\biggr\}.
\label{2.11}
\end{eqnarray}

The limits of the integration over $\tau$ can be extended to
$\pm\infty$.
Then one is left with two integrals. 
The first one equals to
\begin{equation}
\int_{-\infty}^{\infty} {\rm d} \tau\, \tau\, 
\exp\left[{\rm i}\, u (x\tau+a\tau^3)\right]
={2 \over{\rm i}\,}\, {\partial \over \partial x}
\int_{0}^{\infty} {\rm d} \tau\, \cos\left[u (x\tau+a\tau^3)\right]
=
-{2\pi \, {\rm i} \over (3au)^{2/3}}\, {\rm Ai}^{\prime}(z).
\label{2.12}
\end{equation}

Here ${\rm Ai}^{\prime}(z)$ is the derivative of the Airy's function,
and the parameter $z = x\, u/(3au)^{1/3}$.

The second integral reads as follows
\begin{equation}
H \equiv
\int_{-\infty}^{\infty} {\rm d} \tau\, 
{\exp\left[{\rm i}\, u (x\tau+a\tau^3)\right] \over \tau}
=
2\, {\rm i}\, 
\int_{0}^{\infty} {{\rm d} v \over v}\, 
\sin\left(z\, v + {v^3 \over 3}\right).
\label{2.13}
\end{equation}
The derivative of $H$ with respect to $z$ reduces to the Airy's
function: ${\rm d} H/{\rm d} z = 2\pi\, {\rm i}\,{\rm Ai}(z)$. Hence
\begin{equation}
H =
- 2\pi\, {\rm i}\, \int_{z}^{\infty} {\rm d} v \, {\rm Ai}(v).
\label{2.14}
\end{equation}
Here the integration constant is chosen to produce $H=0$ for $z=\infty$.

Substituting (\ref{2.12}) and (\ref{2.14}) into (\ref{2.11}) and
changing the variable of integration from $u$ to $z$ and, afterwards,
integrating by parts the term containing the integral 
$\int_{z}^{\infty} {\rm d} v \, {\rm Ai}(v)$ one gets
\begin{equation}
{ \Delta\, E \over \varepsilon} = 
- {3 \over 2}\, {\alpha \, q^2 \over c} \,L \,
{\varepsilon \over \hbar \gamma^2}\, \chi^2\,
\int_0^{\infty} {z {\rm d} z \over \beta^2(z)}\,  
\left[ 
{ 1+\beta^2(z) \over \beta^2(z)}\, {\rm Ai}^{\prime}(z)
+
{ z^2 \over 3}\, {\rm Ai}(z)
\right],
\label{2.15}
\end{equation}
where $\beta(z)=1 + z^{3/2}\, \chi$.

The parameter $\chi$, the value of which plays the crucial role in
defining the magnitude of the  energy losses, is defined as 
\begin{equation}
\chi =
{\hbar \gamma^3 \over \varepsilon}\, 
\left[ 
{\xi_u^2 \Omega_{u}^2 + \xi_c^2 \Omega_{c}^2 \over 2}
\right]^{1/2}.
\label{2.16}
\end{equation}

To clarify the meaning of $\chi$ let us for a moment ``switch off''
the channeling motion by putting $\xi_{c}=0$. 
Then (omitting the factor $\sqrt{2}$) 
$\chi \approx \hbar \gamma^3 \xi_u \Omega_{u}/\varepsilon =
\hbar \gamma^2 p_u \Omega_{u}/\varepsilon$. 
Here the quantity $\hbar \gamma^2 p_u \Omega_{u}$ is the 
frequency of the radiated intensity maximum, $\omega_{max}$ 
(in the case $p_u>1$)\citeup{Baier}.
For $\omega \gg \omega_{max}$ the intensity  
${\rm d} E/{\rm d} \omega$ exponentially decreases. 
If $\chi < 1$ then the intensity reaches its maximum in  the
``physical'' domain, i.e. $\hbar \omega_{max} < \varepsilon$. 
The opposite case $\chi > 1$ (and, consequently 
$\hbar \omega_{max} > \varepsilon$) corresponds 
to the situation when a projectile can emit photons of all  
frequencies lying within the range $\hbar \omega = [0, \varepsilon]$ 
so that the spectrum intensity never reaches the maximum. 
In this case the radiative energy losses are dominated by the
radiation of highly energetic photons,  $\hbar \omega \sim
\varepsilon$.

The analogous arguments can be provided to analyze the case of the
channeling radiation only, i.e. $\xi_{u}=0$. 

Hence, the formula (\ref{2.16}) is a generalization of the definition
of $\chi$ to the case when both motions, the undulator and the
channeling ones, exist. 
This expression does not contain the cross-term 
$\propto \xi_{u}\,\xi_{c}$ which could have been originated from the 
interference of the photons emitted due to two different types of motion. 
The absence of the cross-term simply reflects the fact that, 
generally,  the frequencies of $\Omega_{u}$ and  $\Omega_{c}$ 
(and, correspondingly, the frequencies of all possible
harmonics) are incompatible and, thus, the corresponding
electromagnetic waves do not interfere.

The expression (\ref{2.15}) can be re-written in the form which is
frequently used in the theory of energy losses due to
the synchrotron and/or undulator radiation\citeup{Baier,Land4}.
To do this let us introduce the energy losses calculated in the
classical limit. 
This limit corresponds to $\hbar \omega/\varepsilon
\ll 1$, which, in turn, means that $\chi \ll 1$.
Neglecting $\chi$ in the integrand in (\ref{2.15}) (it results in
putting $\beta(z)=1$)
one gets:

\begin{equation}
\left({ \Delta\, E \over \varepsilon}\right)_{cl} = 
- {3 \over 2}\, {\alpha \, q^2 \over c} \,L \,
{\varepsilon \over \hbar \gamma^2}\, \chi^2\,
\int_0^{\infty} z {\rm d} z 
\left[ 
2 - 
{ z^3 \over 12}\,\right] {\rm Ai}^{\prime}(z).
\label{2.17}
\end{equation}
Here the second term in the integrand was obtained by integrating by
parts the term proportional to ${\rm Ai}(z)$ in (\ref{2.15}).

The integrals are evaluated by using the formula\citeup{Land4}:
\begin{equation}
\int_0^{\infty}{\rm d} z\,  z^{\nu}\,
{\rm Ai}^{\prime}(z)
=
- {3^{(4\nu-1)/6} \over 2\pi} \, 
\Gamma\left({\nu \over 3} +1\right)\,
\Gamma\left({\nu \over 3} +{1\over 3}\right),
\label{2.18}
\end{equation}
yielding
\begin{equation}
\left({ \Delta\, E \over \varepsilon}\right)_{cl} = 
{2 \over 3}\, {\alpha \, q^2 \over c} \,L \,
{\varepsilon \over \hbar \gamma^2}\, \chi^2.
\label{2.19}
\end{equation}

The functional dependence presented by (\ref{2.19}) coincides with the
well-known expression\citeup{Land4}. 

For a projectile positron (which is, actually, of a prime interest) 
it is convenient to denote the coefficients in (\ref{2.19}) in 
another form by making use of the relations $\varepsilon/(\hbar\,
c\gamma) = m_e c^2/(\hbar\, c) = \alpha/r_e$. Hence
\begin{equation}
\left({ \Delta\, E \over \varepsilon}\right)_{cl} = 
{2 \over 3}\, {\alpha^2 \over r_e} \,L \,
{\chi^2 \over \gamma}.
\label{2.20}
\end{equation}

Intserting  (\ref{2.19}) into (\ref{2.15}) we obtain the formula for the
energy losses within the framework of the quasi-classical approach:
\begin{equation}
{ \Delta\, E \over \varepsilon} = 
\left({ \Delta\, E \over \varepsilon}\right)_{cl} \,
\Phi(\chi), 
\label{2.21}
\end{equation}
where $\Phi(\chi)$ is given by
\begin{equation}
\Phi(\chi) 
=
-{9 \over 4}\, 
\int_0^{\infty} {z {\rm d} z \over \beta^2(z)}\,  
\left[ 
{ 1+\beta^2(z) \over \beta^2(z)}\, {\rm Ai}^{\prime}(z)
+
{ z^2 \over 3}\, {\rm Ai}(z)
\right].
\label{2.22}
\end{equation}

\begin{figure}[htbp]
\epsfig{file=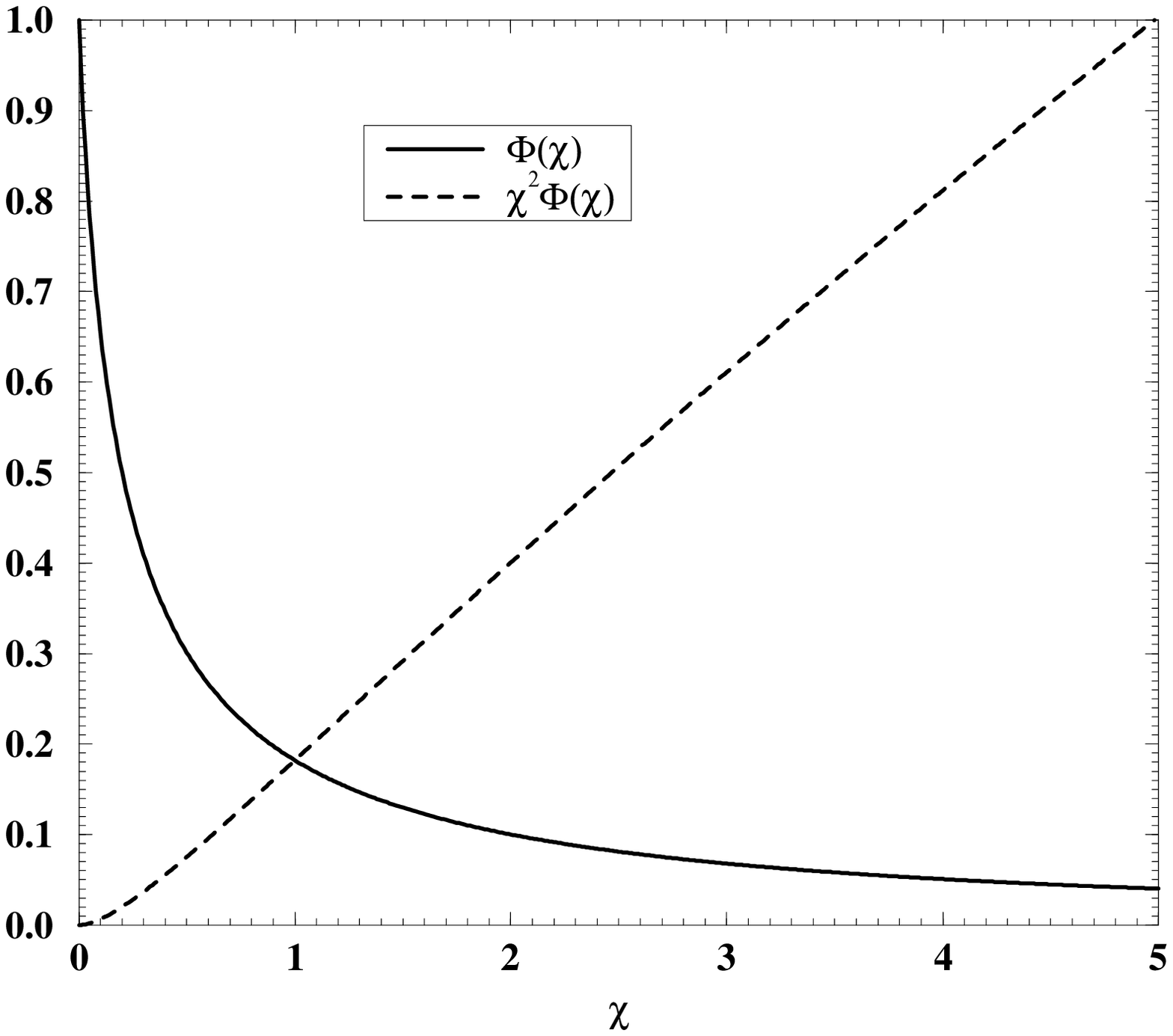,width=11cm, angle=0}
\fcaption{Dependences $\Phi(\chi)$ and $\chi^2\Phi(\chi)$}
\label{fig2}
\end{figure}

The functions $\Phi(\chi)$ and $\chi^2\Phi(\chi)$ are 
presented in Fig. \ref{fig2}.

\subsection{Comparison of the contributions of 
undulator and channeling mechanisms to the energy radiative
losses.}

From (\ref{2.19}) (or (\ref{2.20})) and (\ref{2.21})--(\ref{2.22}) it is
clear that the magnitude of the energy losses for a given projectile
energy and a crystal length  depends solely on the value of the
quantity $\chi$, see (\ref{2.16}). 
Thus, to compare the relative contributions of the two types of
radiation it is sufficient to analyze the ratio 
\begin{equation}
\eta \equiv {\xi_u\, \Omega_u \over \xi_c\, \Omega_c}.
\label{2.23}
\end{equation}
By making use of (\ref{2.4}) and recalling the
definition
$C = \varepsilon/(R_{min}\, q e U^{\prime}_{max})$ 
 one gets
\begin{equation}
\eta = C\,  {d \over |a_c|}.
\label{2.24}
\end{equation}

In a linear channel ($\lambda_u=\infty$, $a_u=0$, $R_{min} =\infty$) 
the range of the $a_c$ values corresponding to the channeling motion 
is $a_c \in [-d/2,d/2]$. 

For a channel  periodically bent by an acoustic wave the range
of the $a_c$ values, for which the stable channeling motion can occur,
is narrower $|a_c| \leq a_c^{(max)}< d/2$. Let us first establish the
magnitude of $a_c^{(max)}$ for given projectile energy and
AW wave length and amplitude.

Inside the channel the motion of the channeled particle
is determined by the effective potential
\begin{equation}
q\, e\, U_{eff}(\tilde{y}) = q\, e\, U(\tilde{y}) - 
{\varepsilon \over R}\, \tilde{y},
\label{2.25}
\end{equation}
where 
$R^{-1}= R^{-1}_{min}\,
\sin(2\pi z/\lambda_u) \in [-R^{-1}_{min}, R^{-1}_{min}]$ 
is the local curvature radius of the channel. 
Written in terms of the dimensionless variable 
$Y =\tilde{y}/d \in \left[-0.5, 0.5\right]$ 
and in the case of the harmonic interplanar potential 
the quantity  $U_{eff}$ reads as
\begin{equation}
U_{eff}(Y) = 
4\, q\, e\, U_o\, \left(Y^2 
- C\,Y\,\sin{2\pi z \over \lambda_u}\right) . 
\label{2.26}
\end{equation}

\begin{figure}[htbp]
\epsfig{file=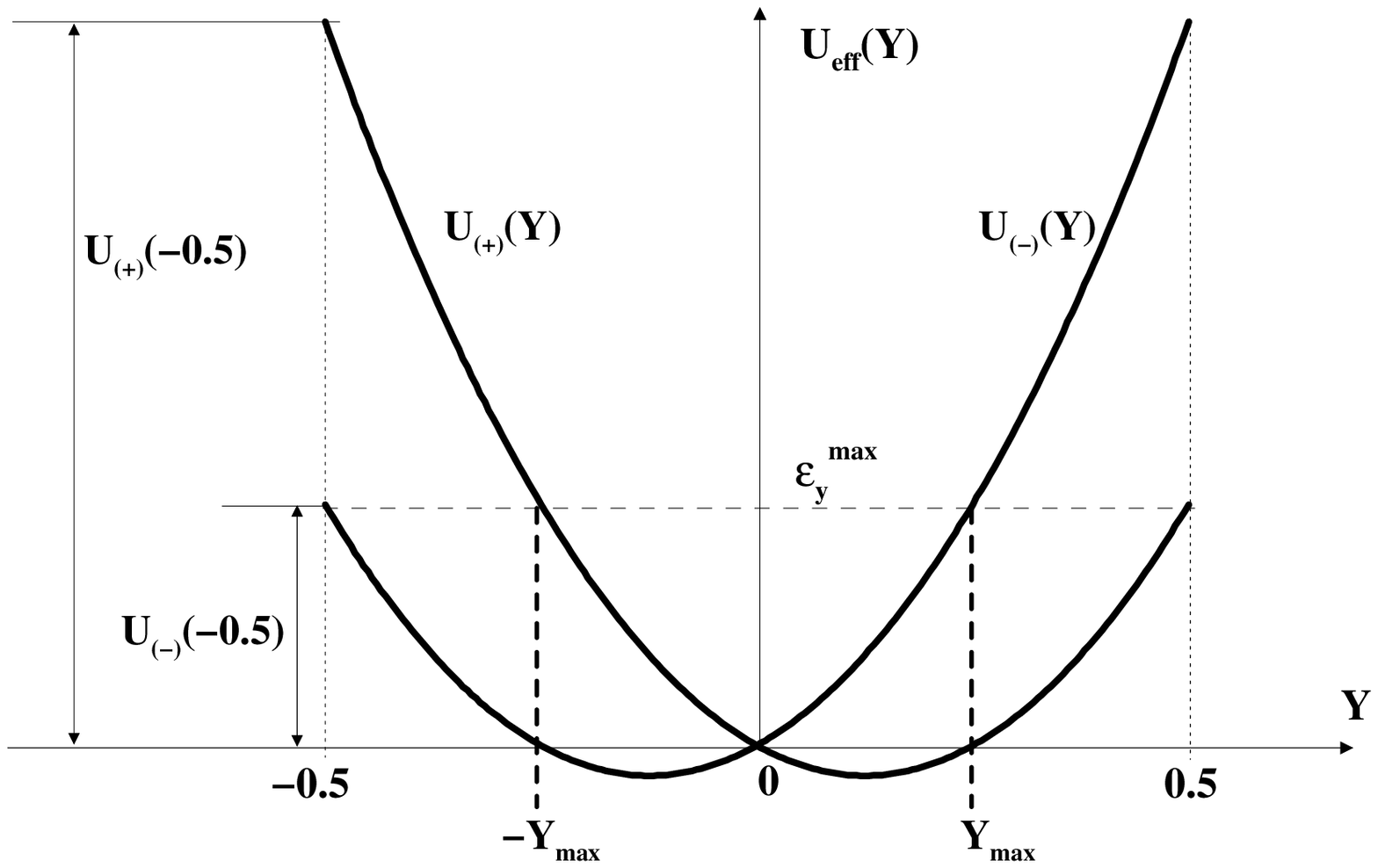,width=11cm, angle=0}
\vspace{0.5cm}
\fcaption{Effective potential $U_{eff}(Y)$ calculated 
in the points of the AW  maximum,  
$U_{(+)}(Y) = 4\, q\, e\, U_o\, (Y^2 - C\,Y)$, and 
minimum,  $U_{(-)}(Y) = 4\, q\, e\, U_o\, (Y^2 + C\,Y)$, (see
(\protect\ref{2.26}))
versus the dimensionless distance from the centerline, 
$Y=2\,\tilde{y}/d \in [-0.5, 0.5]$.
 $\varepsilon_y^{max}=U_{(-)}(-0.5)=U_{(+)}(0.5)$ is the maximum 
transverse energy for which channeling in an acoustically bent crystal
can occur.
The quantity $Y_{max}$ defines the maximum value of the parameter
$a_c$, $Y_{max} = 2 a_c^{(max)}/d$. 
Further explanations are given in the text.}
\label{fig3}
\end{figure}

Schematically, the dependences $U_{eff}(Y)$ calculated 
in the points of the AW maximum ($\sin(2\pi z/\lambda_u) = + 1$) 
and minimum ($\sin(2\pi z/\lambda_u) = - 1$) are presented in
Fig.\  \ref{fig3}. 
The particle will be trapped into the channeling mode only if
its energy, $\varepsilon_{y}$, associated with the transverse motion 
is less than 
$4\, q\,e\, U_o\,\left(|Y|^2-C\,|Y|\right)_{|Y|=0.5}$
i.e. the minimum of two values
$4\,q\,e\, U_o\,\left(|Y|^2\pm C\,|Y|\right)_{|Y|=0.5}$
which are the heights of the asymmetric wells corresponding to the 
effective potential (\ref{2.26}) in the vicinity of the AW minima/maxima.
Hence, in an acoustically bent channel the range $|a_c| \leq a_c^{(max)}$
is determined by  following two inequalities (see Fig.\  \ref{fig3}):
\begin{equation}
Y^2 - C\, Y \leq {1 \over 4}\, (1-C)^2, 
\qquad
Y^2 + C\, Y \leq {1 \over 4}\, (1-C)^2,
\label{2.27b}   
\end{equation}
which result in
\begin{equation}
|a_c| \leq a_c^{(max)} = {d \over 2}\, (1-C).
\label{2.28}
\end{equation}

Let us assume that a uniform beam is ideally collimated along the
centerline when entering the crystal, so that the particles of the beam
differ only in the value of the initial coordinate $\tilde{y}_0$. Then,
the parameter $2\, a_c^{(max)}/d= 1-C$ defines the relative part of
the beam particles which are trapped into the channeling mode in
the acoustically bent crystal.

Equations (\ref{2.24}) and (\ref{2.28}) allow to define the ranges of
the parameter $a_c$ for which either the undulator or the channeling
type of radiation dominate in the total energy losses. 
Namely, for $\eta>1$, and, consequently, $|a_c|<d\, C$,  the undulator
radiation contributes more to the total energy loss. 
If  $\eta>1 \Longrightarrow |a_c|>d\, C$ then the channeling radiation 
plays the dominant role.
If $|a_c|=C$ then both mechanisms contribute equally to the radiative
losses. 
The value $a_c=C\leq a_c^{(max)}$ can be reached  only if $C\leq 1/3$. 
In the opposite case, $C > 1/3$, the parameter $\eta$ is greater than
1 for all $a_c$ values consistent with (\ref{2.28}), and, hence, the
losses due to the undulator radiation are higher than those due to the
channeling one.

\subsection{Estimation of the magnitude of the parameter $\chi$}
 
With (\ref{2.24}) taken into account 
the parameter $\chi$ from  (\ref{2.16}) becomes:
\begin{equation}
\chi = {1 \over \sqrt{2}} \,
\chi_c^{(max)}\, \sqrt{4 C^2 + 4\,{a_c^2 \over d^2}},
\label{2.29}
\end{equation}
where (see (\ref{2.2}) and (\ref{app.A18a}) )
\begin{equation}
\chi_c^{(max)} = 
 {\hbar \gamma^3 \over \varepsilon}\, 
\left[\xi_c \Omega_{c}\right]_{a_c=d/2} =
{\hbar \gamma^3 \over \varepsilon}\, 
{2\, \mu^2 c\over d}
\label{2.30}
\end{equation}
is the maximum value of $\chi$ due to the channeling radiation in the
case of a linear channel, i.e. when $a_c^{(max)} = d/2$. 

The factor $2\, \sqrt{C^2 + a_c^2/d^2}$ in 
(\ref{2.29}) reaches its maximum value of $2$ for $C=1$ (to
get this one substitutes $a_c$ with its maximum value given by 
(\ref{2.28}) and, afterwards, finds the maximum value with respect to
$C$ in the range $C=[0,1]$ within which the channeling in an
acoustically bent crystal can occur).
Hence  
\begin{equation}
\chi_{max} = \sqrt{2}\, \chi_c^{(max)} =
\sqrt{2}\, {\hbar \gamma^3 \over \varepsilon}\, 
{2\, \mu^2 c\over d} .
\label{2.31}
\end{equation}

\begin{figure}[htbp]
\epsfig{file=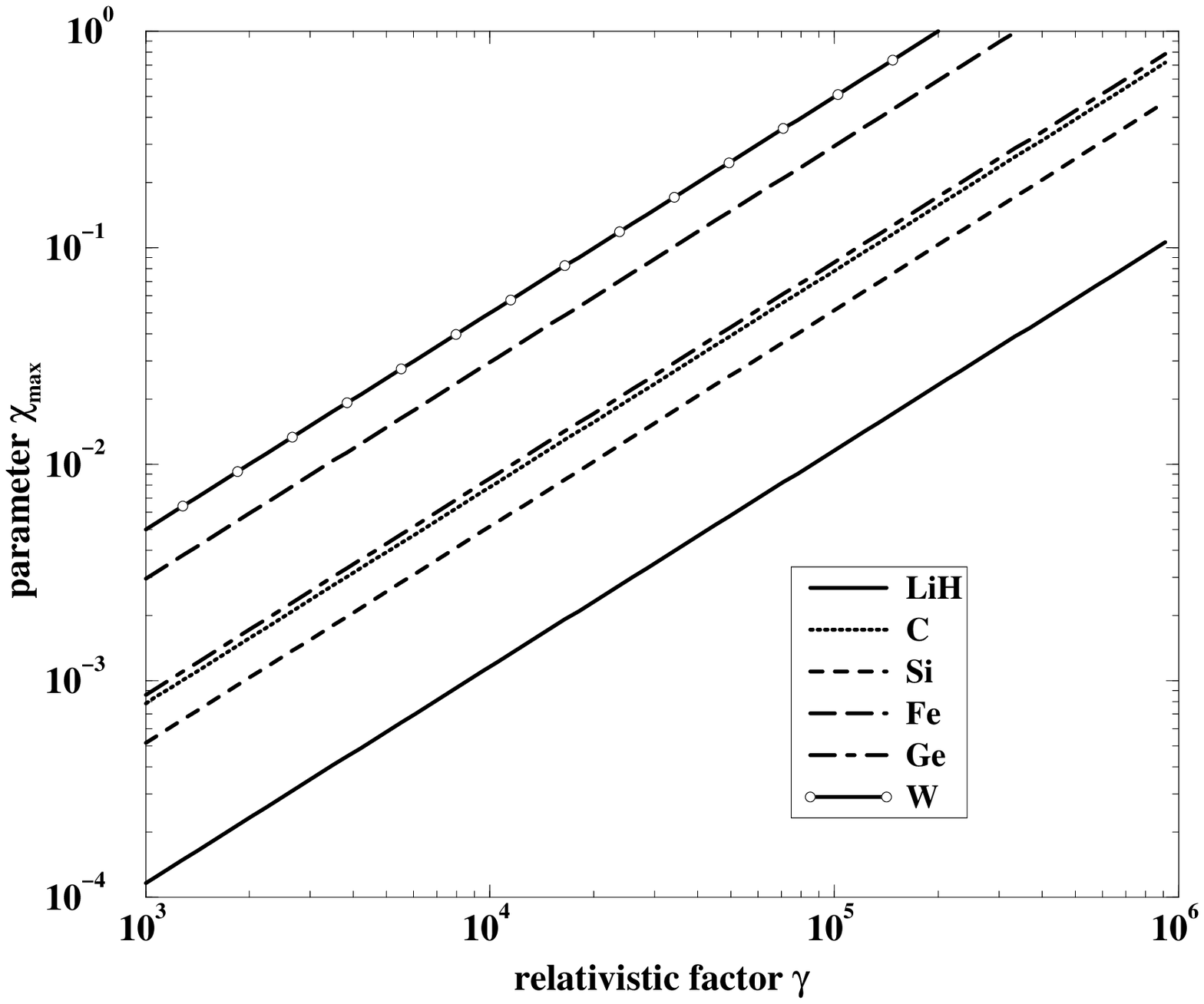,width=11cm, angle=0}
\fcaption{Parameter $\chi_{max}$ from
(\protect\ref{2.32}) versus the relativistic factor of a positron
channeled in (110) channels of various crystals as indicated.}
\label{fig4}
\end{figure}

Let us estimate the magnitude of the right-hand side of (\ref{2.31})
for a positron, $\varepsilon = \gamma\, 0.511\cdot 10^{6}$ eV.
By taking into account the definition (\ref{app.A18a}) one gets after some
simple algebra
\begin{equation}
\chi_{max} = 
4.3\times 10^{-8} \, \gamma\, { v_o \over d_{\AA}}. 
\label{2.32}
\end{equation}
Here $v_o$ is the magnitude of $q\, e\, U_o$ measured in eV and
$d_{\AA}$ is the interplanar spacing measured in \AA.
In table 1 the values of $v_o$ and $d_{\AA}$  
are presented for (110) channels in various crystals as indicated.
The data for C, Si, Fe, Ge and W were taken from Ref. \cite{Baier}, the 
$d_{\AA}$ and $v_o$ values for LiH were adopted from Ref. \cite{LiH}.
The last column of the table corresponds to the values of 
a positron relativistic factor which produce $\chi_{max} = 1$.
Fig.\  \ref{fig4} presents the dependences $\chi_{max}(\gamma)$ 
obtained for the (110) channels of LiH, C, Si,
Fe, Ge and W crystals.

\begin{table}[htbp]
\tcaption{The values of $d_{\AA}$, $v_o$ 
and  $\gamma_{\rm min}$ calculated for a positron channelling in (110) 
planar channel in C, Si, Ge, F and W crystals.}
\centerline{\footnotesize\smalllineskip
\begin{tabular}{rrrr}\\
\hline
Crystal& $d_{\AA}$ & $v_o$ & $\gamma$  for             \\
       &  (\AA)    &  (eV) &  ${\chi_{max} = 1}$\\
\hline
   C   &  1.26     &  23   &  $1.27\times 10^{6}$ \\
  Si   &  1.92     &  23   &  $1.94\times 10^{6}$ \\
  Ge   &  2.00     &  40   &  $1.16\times 10^{6}$ \\
   F   &  1.02     &  70   &  $3.39\times 10^{5}$ \\
   W   &  1.12     &  130  &  $2.00\times 10^{5}$ \\
\hline
\end{tabular}}
\label{Table1}
\end{table}

One may conclude, based on the values of $\gamma_{\chi_{max} = 1}$,
and on the curves in Fig.\  \ref{fig4} that in the range $\gamma < 10^{5}$ 
the parameter $\chi$ can be chosen to satisfy the condition 
$\chi < \chi_{max}<1$.

\subsection{Realistic calculation of the total energy losses by a
positron bunch channeling in an acoustically bent crystal.}

Formulae (\ref{2.20})--(\ref{2.22}) allow to calculate the energy loss
for the particular trajectory (which is specified by the parameter
$a_c$) and for arbitrary crystal length $L$.
A more realistic approach must take into account, firstly, the
effect of the decrease in the beam volume density with the penetration
distance, i.e. the dechanneling effect, and, secondly,  
  the distribution of the beam particles in $a_c$.

\subsubsection{Account for the dechanneling effect.}

Random scattering of the channeling particle by the electrons and nuclei
of the crystal leads to a gradual increase of the particle energy
associated with the transverse oscillations in the channel. As a
result, the transverse energy at some distance from the
entrance point exceeds the depth of the interplanar potential well,
and the particle leaves the channel. 
This distance is called the dechanneling length $L_d(\gamma, R)$. 
For a given crystal and channel it depends on a positron energy 
(relativistic factor) and on the curvature radius $R$

Therefore, for realistic estimations it is sufficient to assume that 
the crystal length $L$ does not exceed $L_d(\gamma, R)$ and
one may calculate
the relative energy loss for $L=L_d(\gamma, R)$.

It can be demonstrated\citeup{Biryukov} that for a crystal bent 
with a constant curvature radius $R$ the dechanneling
length $L_d(\gamma, R)$ satisfies the relation 
\begin{equation}
L_d(\gamma, R) 
= \left(1 - {R_c \over R}\right)^2 L_d(\gamma, \infty),
\label{2.34}
\end{equation}
where $L_d(\gamma, \infty)$ is the dechanneling length of a positron
of the same energy in a straight channel ($R=\infty$) and
\begin{equation}
R_c = {\varepsilon \over e\, q\, U_{\rm max}^{\prime} }
\label{2.35}
\end{equation}
is the critical (minimal) radius consistent with the channeling
condition in a bent crystal, 
``the centrifugal force $<$ the interplanar force''\citeup{Tsyganov}.

In an acoustically bent crystal the curvature 
$R^{-1}(z) = R_{min}^{-1}\,\sin(2\pi z/\lambda_u)$ is not constant.
Therefore, it is natural to consider the mean
curvature $1/\bar{R}$ which is obtained by averaging
$1/|R(z)|$  over the undulator period
\begin{equation}
{1 \over \bar{R}} = {1 \over \lambda_u} \int_0^{\lambda_u}
k_u^2 a_u |\sin k_u z| {\rm d} z = {2  \over \pi}\, 
{1 \over R_{\rm min}}.
\label{2.36}
\end{equation}
Then the dechanneling length in an acoustically bent channel is
estimated as follows:
\begin{equation}
L_d(\gamma, R) 
= (1 - {2  \over \pi}\, C)^2\, \gamma\, \alpha_d\,(\gamma),
\label{2.37}
\end{equation}
where we introduced the reduced dechanneling length 
$\alpha_d (\gamma) \equiv L_d(\gamma, \infty)/\gamma$.
For a given crystal and crystallographic plane this quantity 
depends weakly  on $\gamma$.
Its explicit expression, calculated by using the Lindhard approximation
for the  potential of a planar channel, reads\citeup{Biryukov}
\begin{equation}
\alpha\,  (\gamma)  = {256 \over 9\pi^2}
\,{ a_{\rm TF} \over r_{\rm cl} }\,
{ d \over \ln\left(2\varepsilon /I\right) - 1 }.
\label{2.38}
\end{equation}
Here $a_{\rm TF}=0.8853 Z_c^{-1/3} a_0$ and $I = 16 Z_c^{0.9}$ eV
are the Thomas-Fermi atomic radius and ionization potential,
respectively. $Z_c$ is the atomic number of the crystal atoms,
and $a_0$ is the Bohr radius.
The dependences $\alpha_d(\gamma)$ for various planar channels can be
found in Ref. \cite{lsr_rev}.

Taking into account the quantities introduced above one obtains the
following expression for the relative energy loss due to the
electromagnetic radiation emitted in a crystal of the length
$L=L_d(\gamma, R)$:
\begin{equation}
{ \Delta\, E \over \varepsilon} = 
2.3\times 10^{-7} \, 
\left(\gamma\, { v_o \over d_{\AA}}\right)^2 \,\alpha_d(\gamma)\,
(1 - {2  \over \pi}\, C)^2\,
\left(C^2 + {a_c^2 \over d^2}\right)\, \Phi(\chi) .
\label{2.39}
\end{equation}
Here the factor $2.3\times 10^{-7} \,\left(\gamma\,v_o/ d_{\AA}\right)^2 $
originates from 
$2\, \alpha^2\, \chi^2 /3 r_e$ (see (\ref{2.20})) 
with the parameter $\chi$ taken in the
form given by (\ref{2.29}) and, in turn,  
$\chi_c^{(max)}=\chi_{max}/ \sqrt{2}$ is calculated from (\ref{2.32}).
 
\subsubsection{Energy losses averaged over the parameter $a_c$.}

When a bunch of positrons  enters the crystal the particles have various
values of the initial coordinate $\tilde{y}_0$ and of the incidence angle
$\theta_0$ between the momentum of the incident particle and the
tangent to the crystal centerline $a_u\, \sin(k_u z)$. 
Let us assume that the bunch is ideally collimated so that all the
particles have $\theta_0=0$ at the entrance. 
Furthermore, let us assume that the particles are uniformly distributed in 
the $\tilde{y}_0$ space. 
Then, when considering the energy losses by those
particles of the bunch which are trapped into the channeling mode of
motion,  it is meaningful
to carry out the averaging of $\Delta\, E/  \varepsilon$
over the $a_c$ values satisfying (\ref{2.28}). 
Thus, the average energy losses are defined as follows:
\begin{equation}
\overline{{ \Delta\, E \over \varepsilon}} = 
2.3\times 10^{-7} \, 
\left(\gamma\, { v_o \over d_{\AA}}\right)^2 \,\alpha_d(\gamma)\,
\left(1 - {2  \over \pi}\, C\right)^2\, 
G\left(C, \chi_c^{(max)}\right),
\label{2.40}
\end{equation}
where the function $G(C, \chi_c^{(max)})$ is given by
\begin{equation}
G\left(C, \chi_c^{(max)}\right) =
{2 \over x_0} \, \int_0^{x_0}
{\rm d} x \,
\left(C^2 + x^2\right)  \Phi(\chi). 
\label{2.41}
\end{equation}
Here $x=a_c/d$, 
$\Phi(\chi)$ is defined in  (\ref{2.22}) and 
$\chi=\sqrt{2}\, \chi_c^{(max)} \, \sqrt{C^2 + x^2}$ according
to (\ref{2.29}). 
The upper limit of integration equals  $x_0=a_c^{(max)}/d = (1-C)/2$.

Although the integration over $x$ can be carried out explicitly, the
final result is rather cumbersome to be reproduced here. 
It is simplified in the case $\chi \ll 1$ when\citeup{Land4}
\begin{equation}
\Phi(\chi) =
1 - {55\, \sqrt{3} \over 16}\,\chi + 48\,\chi^2 \, \dots
\label{2.42}
\end{equation}
Then
\begin{equation}
G\left(C, \chi_c^{(max)}\right) =
\left(C^2 + {x_0^2 \over 3} \right)
-
{55\, \sqrt{3} \over 16}\,\chi_c^{(max)} \, A_1 
+
48\,\left(\chi_c^{(max)}\right)^2 \, A_2,
\label{2.42a} 
\end{equation}
where
\begin{eqnarray}
 A_1 &=& 
{1 \over 8}\,
\left\{
\left(C^2 + x_0^2 \right)^{1/2}\left[5 C^2 + 2 x_0^2 \right]
+
{3C^4 \over x_0}\, {\rm ln}{x_0 + \sqrt{C^2 + x_0^2} \over C}
\right\},
\label{2.42b} \\
 A_2 &=& C^4 + {2 \over 3}\,C^2 x_0^2 + {1 \over 5}\,x_0^4.
\label{2.42c}
\end{eqnarray}

\subsubsection{Results of the numerical calculations of the averaged
energy losses.}

As it was mentioned in Sec.\, \ref{Introduction} there are two types
of radiation accompanying the channeling process of an ultra-relativistic
particle in an acoustically bent channel: the ordinary channeling
radiation and the AIR. Both of them belong to an undulator type 
of radiation.    
It is known from general theory of a planar undulator radiation 
(see e.g. Ref. \cite{Kniga}) that its frequency-angular distribution 
${\rm d}E_{\omega}({\bf n})/{\rm d}\omega{\rm d}\Omega_{\bf n}$
is represented by the sets of 
characteristic frequencies (harmonics) 
$\omega^{(K)}$ ($K=1,2,3 \dots$) each of the width 
$\Gamma^{(K)} = (2/N)\,(\omega^{(K)}/K)$, where 
$N$ is the number of the undulator periods.  
The frequencies of
harmonics are defined from the relation
$\hbar\, \omega^{(K)} = 4 \gamma^2 \Omega \, K/(2+p^2)$ with $\Omega$ and
$p$ standing, respectively,  for the undulator frequency and
parameter.
 
In the present paper we analyze mainly the case 
$\Omega_c/\Omega_u \gg 1$ (see Eq.\, (\ref{2.00})) which can be
achieved by choosing $a_u \gg d$ (see Eq.\, (\ref{app.A21})).
Hence, the frequencies of harmonics characterizing the AIR and 
the channeling radiation are well separated satisfying the condition 
$\omega_c^{(K)}/\omega_u^{(K)} \gg 1$ (Refs. \cite{lsr_rev,our_new}).
It is the AIR mechanism which brings the novelty into the problem. 
Therefore, let us analyze the stability of the undulator AIR
radiation towards the decrease in a projectile positron energy due to
the radiative losses.

The frequencies of harmonics of the AIR radiation one calculates from 
(see  Ref. \cite{lsr_rev})
\begin{equation}
\hbar\, \omega_u^{(K)} 
= {4 \gamma^2 \Omega_u\, K \over 2+2\theta^2\gamma^{2}+ p_u^2},
\label{2.43a}
\end{equation}
where $p_u = \gamma \xi_u$, and $\theta$ is the emission angle 
with respect to the undulator axis (which is the centerline of 
the initially linear channel, see Fig.\, \ref{fig1}). The magnitude of
$\theta$ satisfies to 
$\theta \leq \theta_{max} = max\{\gamma^{-1},\xi_u\}$ (Ref. \cite{lsr_rev}). 

From  (\ref{2.43a}) one gets the following relation between the total 
energy losses $\Delta\, E / \varepsilon$ and the quantity 
$\Delta \omega_u^{(K)}$ which is the shift of the $K$th harmonic 
frequency from its unperturbed value $\omega_u^{(K)}$
\begin{equation}
\Delta \omega_u^{(K)} 
= \omega_u^{(K)}\, {2 \over 2+2\theta^2\gamma^{2}+ p_u^2}\,
{2\, \Delta E \over \varepsilon}
< \omega_u^{(K)}\,{2\, \Delta E \over \varepsilon}
\label{2.43b}
\end{equation}

The spontaneous AIR radiation formed during the passage of a 
positron through an acoustically bent crystal
of total length $L_d(\gamma, R)$ is stable towards the energy loss of
the positron provided the shift $\Delta \omega_u^{(K)}$ is smaller
than the natural line half-width $\Gamma_u^{(K)}/2$. 
The latter is given by
$\Gamma_u^{(K)}/2 = (1/N_u)\,(\omega_u^{(K)}/K)$, 
where  $N_u=L_d(\gamma, R)/\lambda_u$.
Therefore, from (\ref{2.43a}) one deduces
\begin{equation}
\overline{{\Delta\, E \over \varepsilon}} \leq {1 \over 2 K N_u}.
\label{2.43c}
\end{equation}

It was estimated in Ref. \cite{lsr_rev} and analyzed in more detail in 
Ref. \cite{our_new} that the realistic range of $N_u$ in an acoustically
based undulator is $N_u = 10 \dots 25$ and the corresponding number of 
the harmonics emitted via the AIR mechanism is $K \sim 1$. 
Thus, the stability of the AIR radiation will occur if 
$\overline{\Delta E}/ \varepsilon < 0.01$.

\begin{figure}[htbp]
\epsfig{file=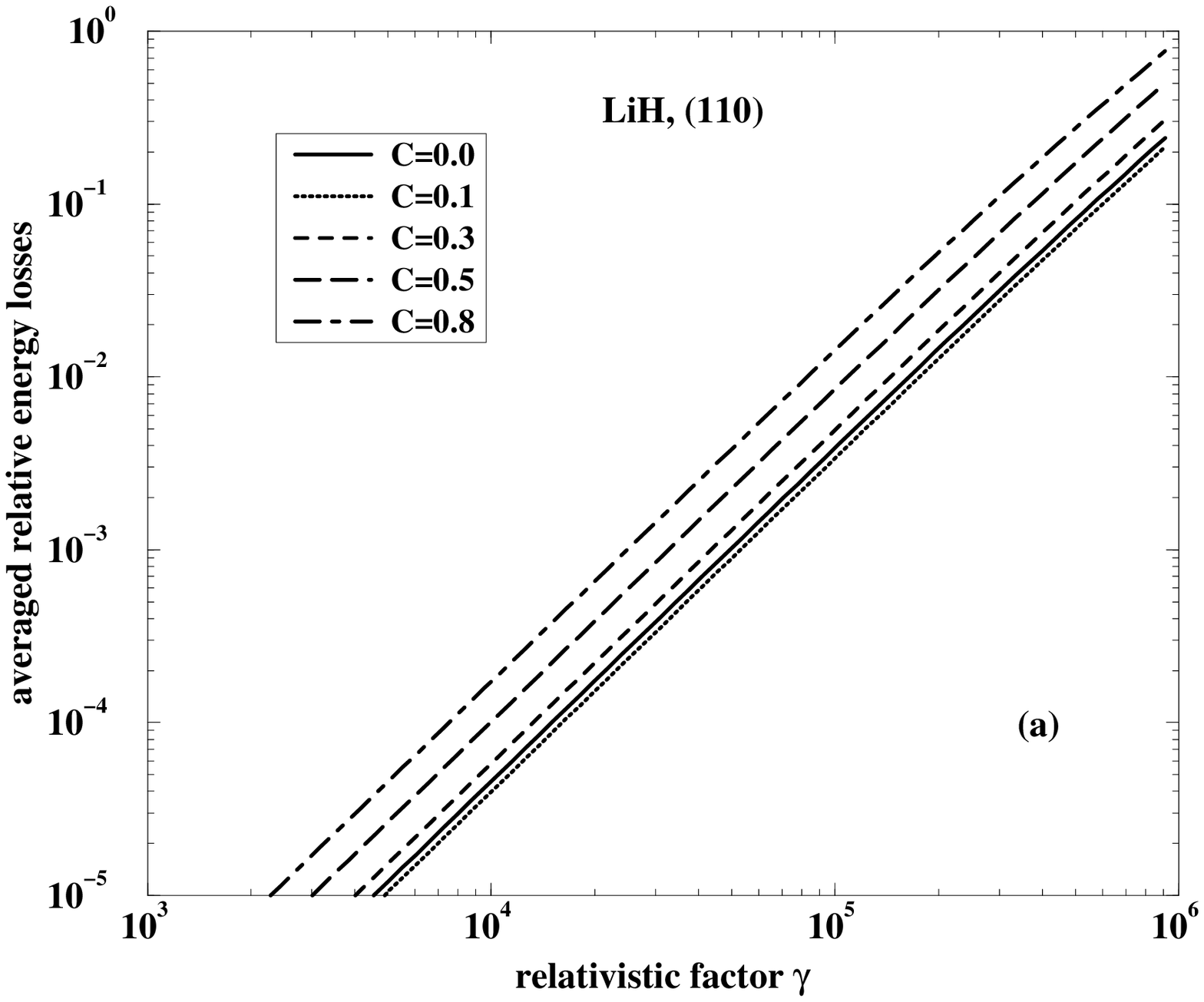,width=11cm, angle=0}
\vspace{1cm}
\epsfig{file=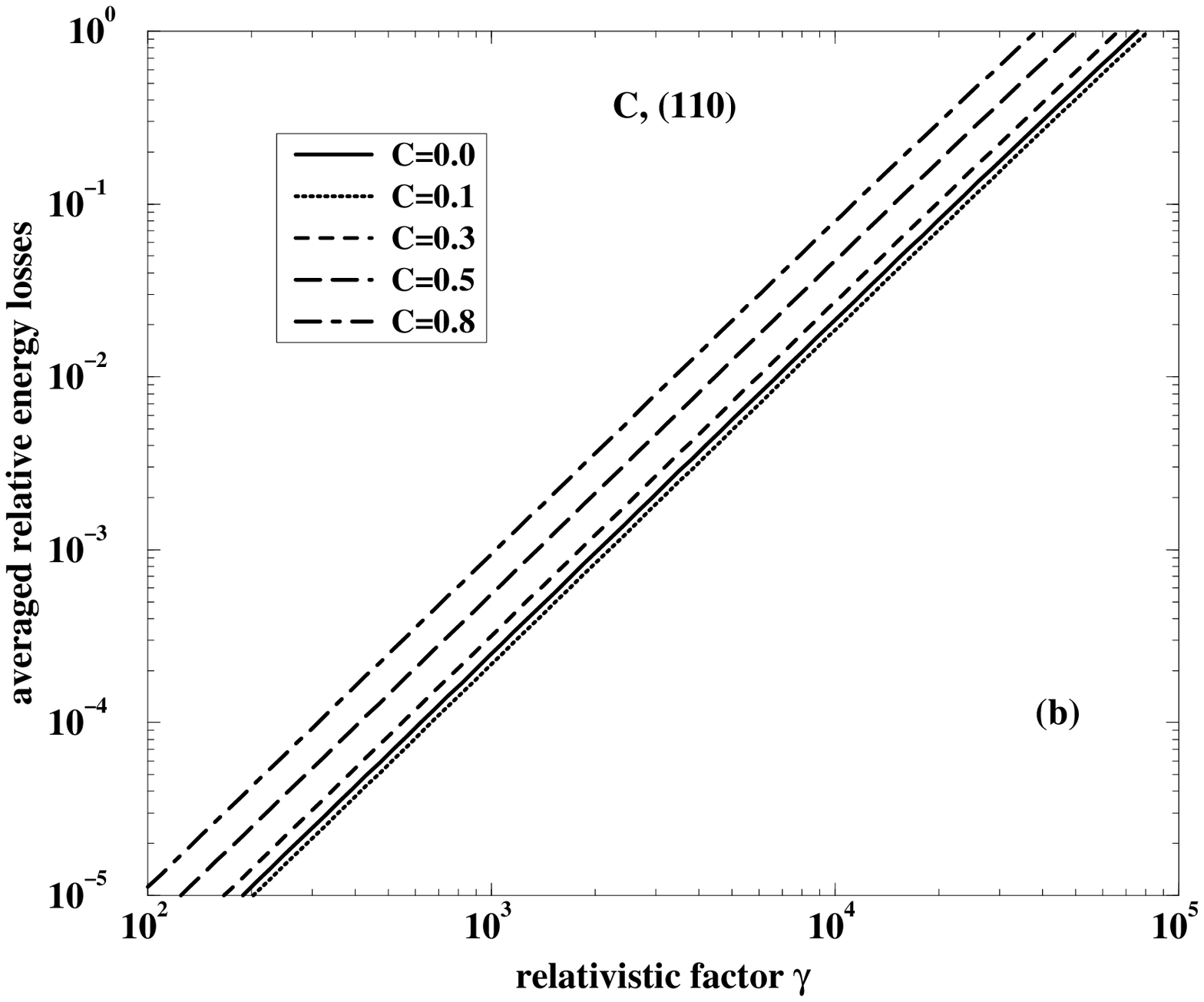,width=11cm, angle=0}
\vspace{0.5cm}
\fcaption{Averaged relative energy losses (\protect\ref{2.40})
for the crystal
length $L=L_d(\gamma,R)$ versus the relativistic
factor of a positron channeling in (110) channels of various crystals:
{\bf (a)} LiH, 
{\bf (b)} C (diamond),
{\bf (c)} Si,  
{\bf (d)} Fe,  
{\bf (e)} Ge,  
{\bf (f)} W.   
The curves corresponds to different values of the parameter 
$C=\varepsilon/(R_{min}\, q e U^{\prime}_{max})$ 
as indicated.}
\label{fig5}
\end{figure}

Figures \ref{fig5} represent the dependences 
$\overline{\Delta\, E/ \varepsilon}$
versus $\gamma$ calculated, according to (\ref{2.40})--(\ref{2.41}), 
for a positron channeling in (110) channels
of LiH, C, Si, Fe, Ge and W crystals and for several values of the
parameter $C<1$ which characterizes the bending of the channel. 

The chosen crystals are commonly used in experiments devoted to the
investigation of the channeling phenomena and, in addition,  
this set includes crystals composed of light (LiH, C, Si), 
intermediate (Fe, Ge) and heavy (W) atoms. 

Figures \ref{fig5}  allow to estimate the range of validity of the 
condition (\ref{2.43c}) for $N_u = 10 \dots 25$.
It is seen from the figures that the inequality (\ref{2.43c}) is
well-fulfilled for $\gamma < 10^5$ in the case of LiH crystal,
$\gamma < 5\times 10^3$ for C, Si, and Ge, 
$\gamma < 2\times 10^3$ for Fe and W.

For higher values of $\gamma$ the increase of 
$\overline{\Delta\, E/ \varepsilon}$ will result in the shift 
$\Delta \omega_u^{(K)}$ greater than the half-width
$\Gamma_u^{(K)}/2$ thus smearing the line over the wider range of
frequencies. In this case it is meanigful to consider not the harmonic 
acoustic wave but rather the one with varying amplitude and period
analogously to how it was proposed when considering undulator
radiation formed in the tampered magnetic wigglers\citeup{KMR}.

\section{Conclusions.}

In this work we have described the general formalism for the
calculation of the total radiative energy loss accounting for the
contributions of both radiation mechanisms, i.e.  the acoustically
induced radiation\citeup{laser,lsr_rev} and the ordinary channeling
radiation.  Our formalism is based on the quasi-classical approach
(see e.g. Ref. \cite{Baier}).  We have analyzed the relative
importance of ordinary channeling radiation and the AIR to the total
radiation energy loss at various amplitudes and lengths of the
acoustic wave and as a function of the energy of the projectile
particle.  We established the ranges of the projectile particle
energy, in which the total radiative energy loss is negligible for the
LiH, C, Si, Ge, Fe and W crystals.  This result is important for the
determination of the projectile particle energy region, in which
acoustically induced radiation of the undulator type and also the
stimulated photon emission can be effectively generated.

We consider our present research as a milestone for the advanced
theoretical description of the AIR phenomenon.  The goal of further
investigation in this field is to achieve an accurate quantitative
description of the undulator radiation and of the corresponding laser
effect.  In our recent work\citeup{lsr_rev} we have outlined the
phenomena, which must be thoroughly considered.

Here we mention only the problem, which is closely connected to the
present research.  Using the formalism very similar to the one
described here, it is interesting to calculate the total frequency and
angular distribution of of photons emitted due to the mechanisms of
the ordinary channeling radiation and the AIR.  This work is in
progress at the moment and will become the subject of another
publication in the near future.

\begin{figure}[htbp]
\epsfig{file=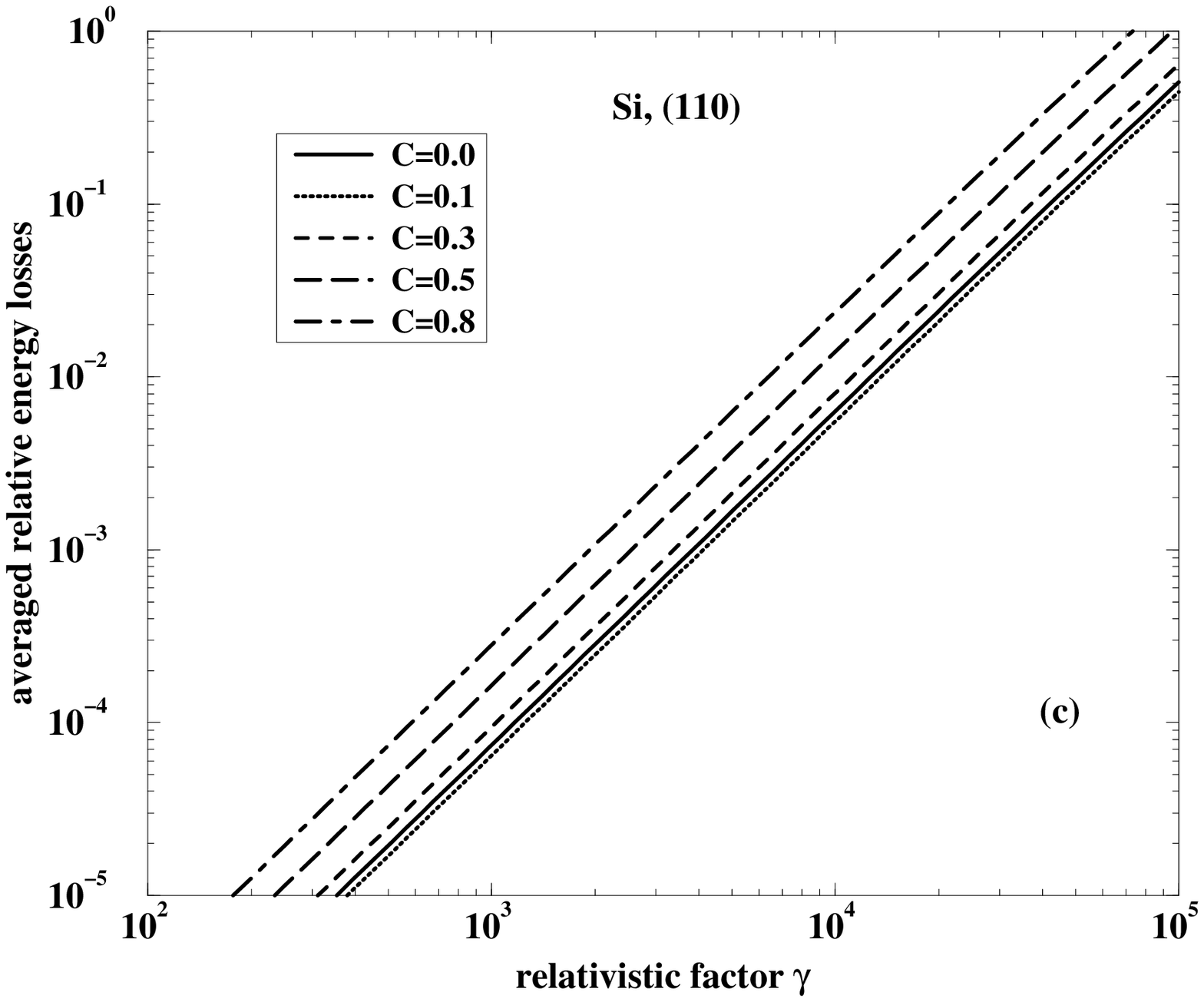,width=11cm, angle=0}
\end{figure}
\begin{figure}[htbp]
\epsfig{file=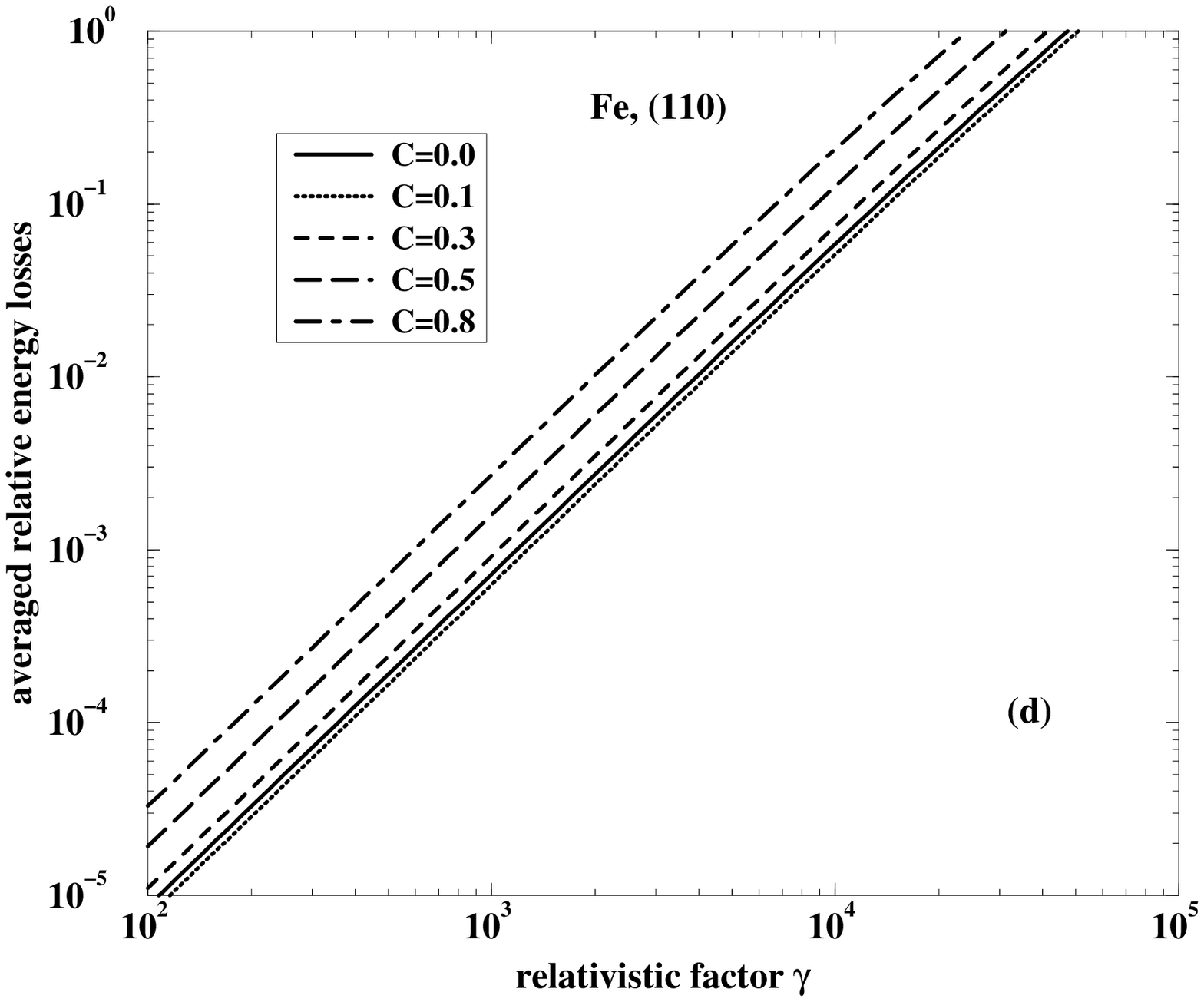,width=11cm, angle=0}
\vspace{0.5cm}
\centerline{Fig. 5 ({\it Continued})}
\end{figure}

\begin{figure}[htbp]
\epsfig{file=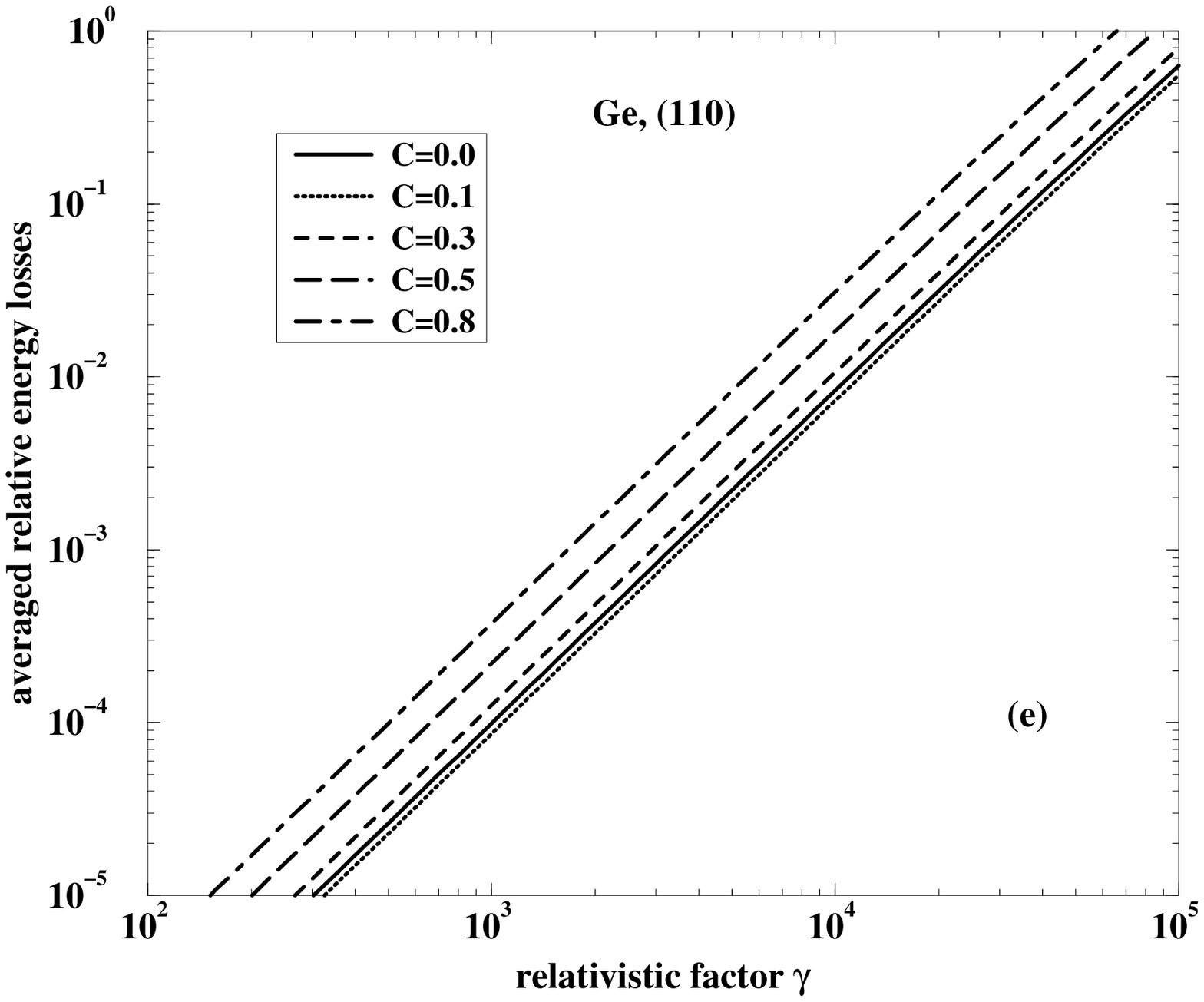,width=11cm, angle=0}
\end{figure}
\begin{figure}[htbp]
\epsfig{file=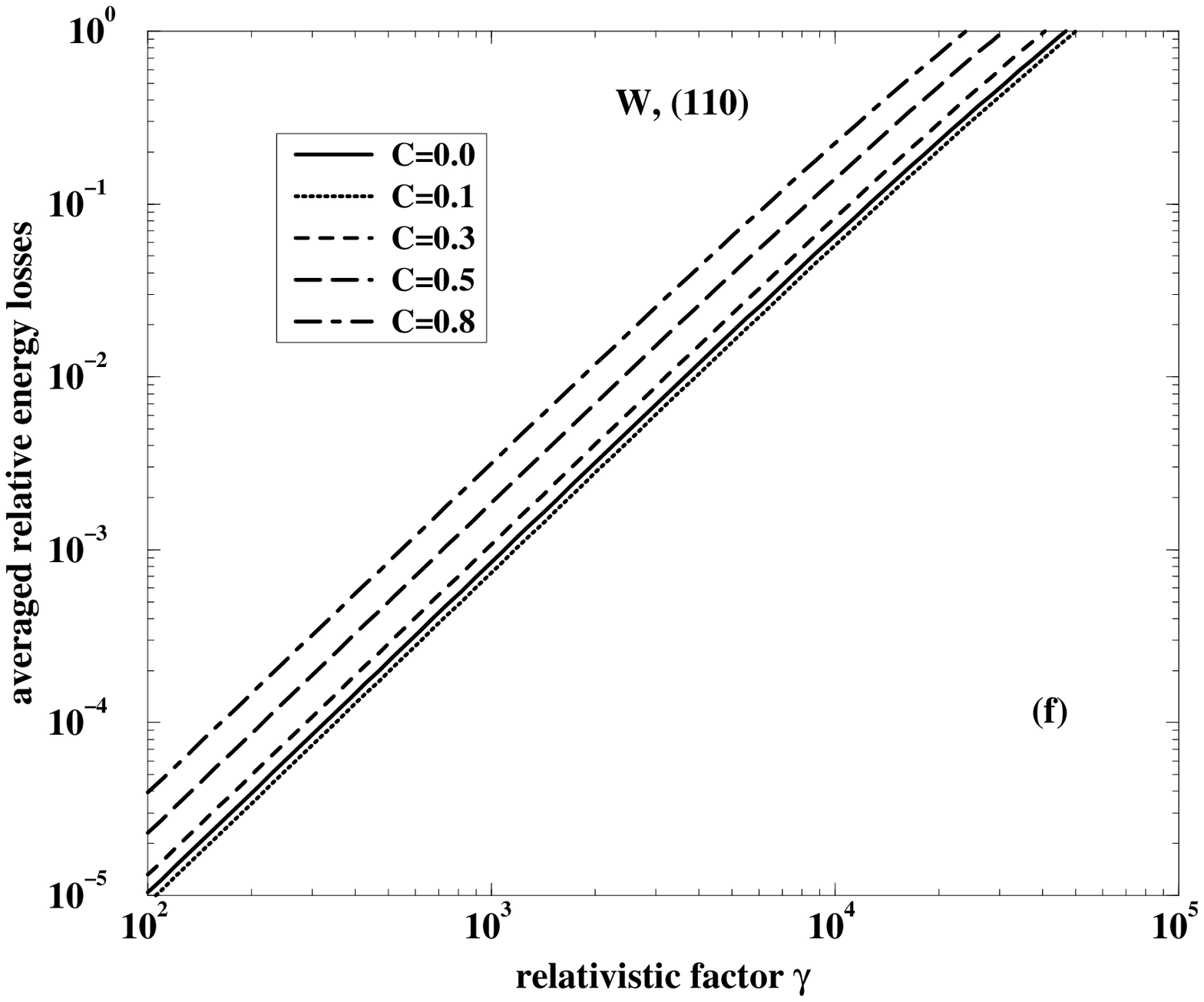,width=11cm, angle=0}
\vspace{0.5cm}
\centerline{Fig. 5 ({\it Continued})}
\end{figure}

\nonumsection{Acknowledgements}
The authors acknowledge support from the DFG, GSI, BMBF and
the Alexander von Humboldt Foundation.

\newpage

\appendix

\setcounter{section}{0}
\renewcommand{\labelenumii}{\Alph{enumi}.\arabic{enumi}}

\section{Particle's motion in an acoustically bent channel}
\label{AppendixA}

\subsection{Approximations}
\label{Approx}

In this section we outline the approximations which have been used
when considering both the motion and the radiation of an
ultra-relativistic charged particle undergoing planar channeling in a
crystal bent by means of a {\it transverse, harmonic, plane} acoustic
wave transmitted along the $z$-direction which coincides with a
crystallographic direction in the initially linear crystal.  It is
assumed that the crystallographic planes are equally spaced and being
parallel to the $(xy)$-plane.

The ultra-relativistic particle enters the crystal at $z=0$ having
only the $y-$ and $z-$ velocity components, $v_{yo}$, $v_{zo}$.  It is
assumed that $v_{yo} \ll v_{zo}$ and $v_{zo}\approx c$.  If one
neglects random scattering of the particle by the electrons and nuclei
of the crystal, then the particle's trajectory lies in the
$(xy)$-plane and is subject to the joint action of the interplanar
force $U^{\prime}(\rho)$ and of the centrifugal force due to the
crystal bending.

The necessary (but not sufficient) condition for a projectile to be
trapped into the channeling mode of motion in a bent crystal is
$\Theta < \Theta_c$, where $\Theta$ is the entrance angle between the
particle's velocity and the channel centerline, and $\Theta_c$ is some
critical angle (the estimates of $\Theta_c$ in the case of bent
channel can be found in Ref. \cite{Biryukov}). In a linear crystal
$\Theta_c$ coincides with the Lindhard's angle\citeup{Gemmell}.

The strong inequality $\Theta \ll \Theta_c$ allows to introduce the
continuum approximation for the interaction potential $U$ between the
charged projectile and lattice atoms arranged in atomic planes.  In
our paper this approximation is used to describe the equations of
motion for the particle in a bent channel.

\subsubsection{Approximations related to the crystal bending}
\label{Approx1}

The shape of a channel centerline in an acoustically bent crystal is 
described by 
\begin{equation}
y(z) = a_u\, \sin k_u z,
\label{app.A1}
\end{equation}
with $k_u = 2\pi/\lambda_u$.

It was demonstrated in Ref. \cite{lsr_rev} that the realistic ranges for
the AW amplitude and wavelength are 
\begin{equation}
 a_u = 10^{-8}\dots  10^{-6}\ {\rm cm}, 
\qquad
  \lambda_u = 10^{-3}\dots  10^{-1}\ {\rm cm}. 
\label{app.A2}
\end{equation}
Therefore, we introduce an approximation by assuming that the
following strong inequality is fulfilled:
\begin{equation}
\xi_u = 2\pi\,  {a_u \over \lambda_u} \ll 1.
\label{app.A3}   
\end{equation}
All the final formulae written below take into account the
contributions of the terms proportional to $\xi_u^{0}$ and 
$\xi_u^{1}$, while the terms the order $\xi_u^{2}$ and higher are
omitted (except for the cases when the $\xi_u^{2}$ terms are the 
leading ones).

\begin{figure}[htbp]
\epsfig{file=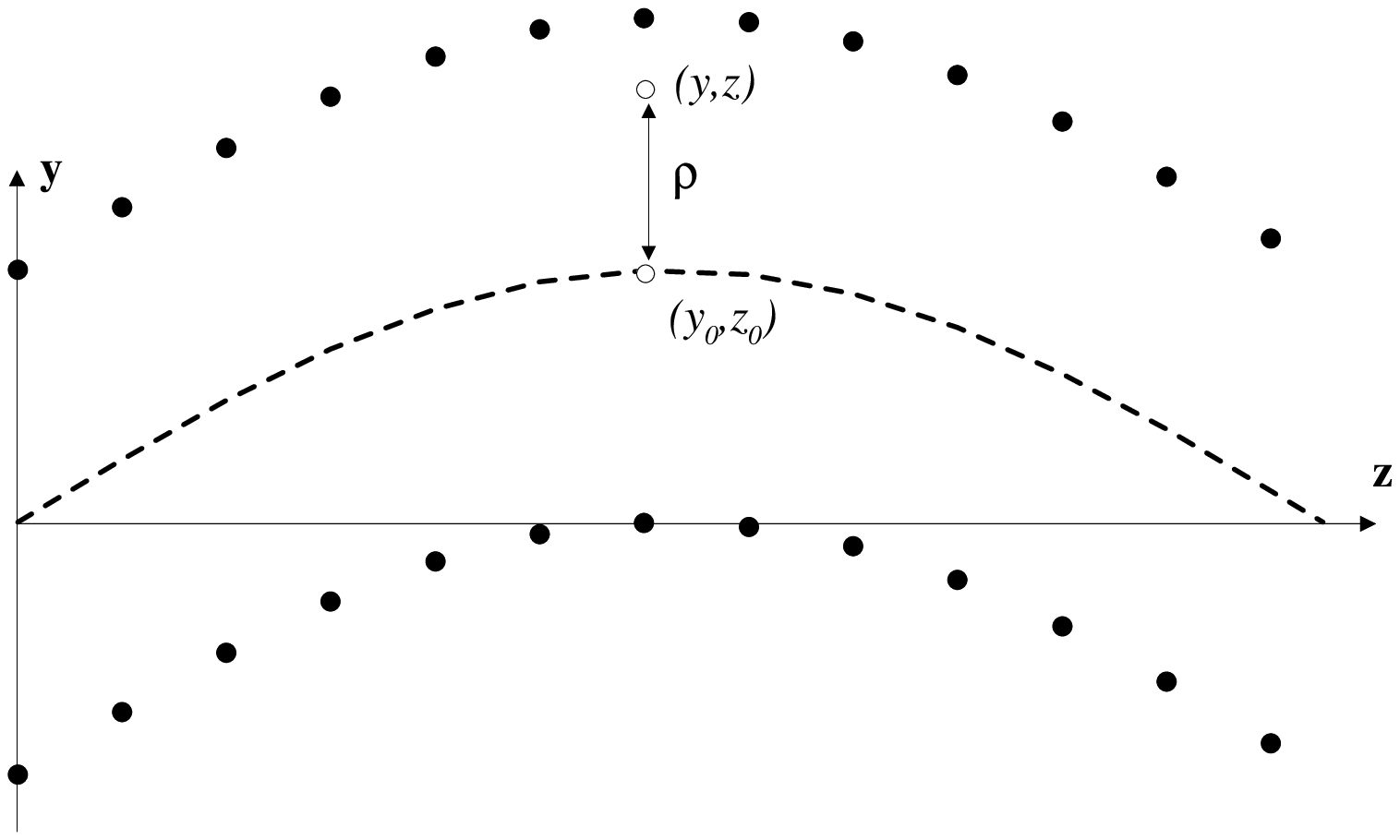,width=12cm, angle=0}
\vspace{0.5cm}
\fcaption{The coordinates used to describe a projectile
position in an acoustically bent channel: $(y,z)$ are the Cartesian
coordinates of the particles, $\rho=\sqrt{(y-y_o)^2 +(z-z_o)^2}\approx
(y-y_o)$ is the distance to the channel centerline.}
\label{fig6}
\end{figure}

It is easily verified that the length of a centerline and
the interplanar spacing in linear ($L, d$) 
and in acoustically bent ($L^{\prime}, d^{\prime}$) channels 
are related as follows  
\begin{equation}
L^{\prime} = L\, (1 + O(\xi_u^2)),
\qquad
d^{\prime} = d\, (1 + O(\xi_u^2)).
\label{app.A4}    
\end{equation}
The analogous relationship one finds for the distance $\rho$ between 
some inner point  $(y,z)$ of the bent channel and its centerline 
(see  Fig. \ref{fig6}):
\begin{equation}
\rho^2  = (y-y_o)^2 + (z-z_o)^2 
= (y-y_o)^2\left( 1+ {(z-z_o)^2 \over (y-y_o)^2}\right)
= (y-y_o)^2\,(1 + O(\xi_u^2)).
\label{app.A5}    
\end{equation}
Therefore, when neglecting the terms of the order $\xi_u^2$ and higher
one may put $L^{\prime}=L$, $d^{\prime}=d$, and disregard the
difference between $\rho$ and $y -a_u\, \sin (k_u z)$.

Eqs. (\ref{app.A4})--(\ref{app.A5}) allow us to introduce the constant
field approximation for the interplanar potential\citeup{Baier} which
assumes that: (1) within any bent channel the potential $U$ depends
only on the variable $\rho$ which is the distance of the $(y,z)$ point
from the channel centerline, (2) the explicit dependence of $U$ on
$\rho$ in the acoustically bent channel is identical to the $U(\rho)$
dependence in the linear channel.

In a linear channel, where the coordinates $y$ and $\rho$ are
basically the same, the potential depends only on the transverse
coordinate $y$ and is a periodic function with the period $d$: $U
\equiv U(y) = U(y+N\, d)$, where the index $N=\dots, -1,0,1,\dots$
enumerates the channels.

When the channel is bent by the AW then the $y$ coordinate becomes $y
= a_u\, \sin (k_u z)$.  The displacement along the $z$ axis is
proportional to $\xi_u^2$ and, thus, may be disregarded.  Hence, the
magnitude of $U$ in the linear channel in the point $(y, z)$
corresponds to the the magnitude of the potential in the bent channel
calculated at the point $(y + a_u\, \sin (k_u z), z)$.  The dependence
of $U$ on $y$ and $z$ in an acoustically bent channel is given by
\begin{equation}
U(y,z) = U(\tilde{y}),
\qquad           
\tilde{y} = y - a_u\, \sin (k_u z)
\label{app.A6} 
\end{equation}
The constant field approximation means that the explicit dependence
$U(\tilde{y})$ in the bent channel is equivalent to the dependence
$U(y)$ in a linear one. The coordinate $\rho$, in the bent channel up
to the terms $\sim \xi_u$, is equivalent to $\tilde{y}$.

\subsubsection{Approximations related to the energy of a projectile}
\label{Approx2}

We assume that for an ultra-relativistic particle, which enters the
crystal having the energy $\varepsilon_0 = m \, c^2\,
\gamma_0$ (here the index $``0''$ indicates that the quantity is
measured at the entrance) the following condition is valid
while it channels in the crystal
\begin{equation}
{q\,e\,  U_o \over \varepsilon_0} \ll 1.
\label{app.A7}
\end{equation}
Here the quantity $U_o$ stands for the depth of the interplanar
potential well.  For a positron ($q=1$) typical values for $q\,e\,
U_o$ are $\sim 10\dots 100$ eV, so that (\ref{app.A7}) is well
fulfilled in the ultra-relativistic case.

The strong inequality (\ref{app.A7}) justifies the classical
description of the particle's motion in both linear and bent crystals.

\subsection{The equations of motion}
\label{motion}

The Hamiltonian function of a relativistic particle moving in a scalar
potential $U$ is given by
\begin{equation}
H = \sqrt{(c\, {\bf p})^2 + m^2 c^4} + q\,e\, U(\tilde{y}),
\label{app.A8}
\end{equation}
where ${\bf p} = m \gamma {\bf v}$ is the momentum, and the coordinate
$\tilde{y}$ is defined in (\ref{app.A6}).

This Hamiltonian does not depend on time. Therefore, the total energy
of the particle, $m\gamma c^2 + q e U$ is conserved.  Hence, the
relativistic factor satisfies the condition $\gamma =\gamma_0\,
\left(1 - q e U/\varepsilon_0 \right)$.  If one neglects the term $q
\,e\, U /\varepsilon_0$ then the relativistic factor
\begin{equation}
 \gamma^{-1} = 
\sqrt{1  - {\dot{y}^2 \over c^2} - {\dot{z}^2 \over c^2}}
\label{app.A9}
\end{equation}
is the integral of motion $\gamma = \gamma_0 = const$.

By introducing the variable $\tilde{y}$ in the equation of motion $
\dot{\bf p} = - \partial H/\partial {\bf r}$ one derives
\begin{eqnarray}
\ddot{\tilde{y}} & =&
- {q e \over m \gamma }\, {{\rm d}  U \over {\rm d}  \tilde{y}}
+
 k_u \xi_u\,c^2\,\sin(k_u c t),
\label{app.A10} \\
\ddot{z}& =& 0.
\label{app.A11}  
\end{eqnarray}
This system was obtained by omitting the terms $\propto q \,e\, U
/\varepsilon_0$ and $\propto \xi_u^n$, $n\geq 2$.  The first term on
the right-hand side of (\ref{app.A10}) represents the acceleration due
to the action of the interplanar force.  The second one is due to the
channel bending and can be written in the form $ \gamma v^2/R$ which
explicitly indicates the centrifugal acceleration (here, $v\approx c$
for an ultra-relativistic projectile, and $R^{-1} = k_u \xi_u\,
\sin(k_u z)$ with $z\approx c t$ is the curvature of the centerline
(\ref{app.A1})).  This term vanishes in the case of a linear channel
when $a_u=0$ and/or $\lambda_u \longrightarrow \infty$.

The equation (\ref{app.A11}) is readily integrated yielding
$z= c t$. 
The correction to this dependence one finds by using 
the relation $\gamma = \gamma_0 = const$ in (\ref{app.A9}): 
\begin{equation}
{v_z^2(t)\over c^2} = 1- 
\left({1 \over \gamma^2} + {v_y^2(t) \over c^2}\right).
\label{app.A12}
\end{equation}
Let us now estimate the ratio $v_y^2(t)/c^2$.  There are two typical
scales for the velocity in the $y$ direction.  The first one,
$v_y^{(1)}$ is related to the motion of the projectile along the
centerline of the channel. The period of this motion equals $\approx
\lambda_u/c$, hence, $v_y^{(1)} \sim 2a_u/(\lambda_u/c)$.  The second
characteristic velocity, $v_y^{(2)}$, is connected with the particle
oscillations inside the channel due to the action of the interplanar
field $U$.  The period of this oscillations is estimated as
$\tau_c\sim 2\pi\sqrt{m\gamma/q e U^{\prime\prime}} \sim
\pi\, d\sqrt{m\gamma/q e U_o}$, so that $v_y^{(2)} \sim 2d/\tau_c$.
Hence
\begin{equation}
\left({v_y^{(1)} \over c }\right)^2 
 \sim \xi_u^2 \ll 1,
\qquad          
\left({v_y^{(2)} \over c }\right)^2 \sim 
{q e U_o \over \varepsilon_0}   \ll 1.
\label{app.A13}
\end{equation}
These estimates, combined with the relation
$v_y(t) = \dot{y} = \dot{\tilde{y}} + \xi_u\,c \cos(kct)$
(see (\ref{app.A1}) and (\ref{app.A6})) produce
\begin{equation}
z(t) = ct + \Delta z(t), 
\label{app.A14}
\end{equation}
with $\Delta z(t)$ satisfying the equation
\begin{equation}
{{\rm d} \Delta z  \over {\rm d} t}
= - {c \over 2 }\, 
\left[{1 \over \gamma^2} + 
{\left(\dot{\tilde{y}}(t) + \xi_u\,c \cos(kct) \right)^2 \over c^2}
\right].  
\label{app.A15}
\end{equation}

For an arbitrary function $U(\tilde{y})$ the system
(\ref{app.A10})--(\ref{app.A15}) can be easily integrated numerically
by setting the initial conditions $\tilde{y}(0),\,
\dot{\tilde{y}}(0),\, \Delta z(0)$.

The function $\tilde{y}(t)$ describes the motion of the particle with
respect to the centerline (\ref{app.A1}) of the acoustically bent
channel.  According to (\ref{app.A6}), the total $y(t)$ dependence is
obtained by combining $\tilde{y}(t)$ and the term $a_u\, \sin (k_u c
t)$.

\subsection
{As a specific case: harmonic approximation for 
the interplanar potential}
\label{harmonic}

The case of the harmonic interplanar potential is of a particular
interest because it allows an analytical solution of the equations of
motion.  Substituting the function (\ref{2.0}) into the right-hand
side of (\ref{app.A10}) one gets the well-known equation for a driven
pendulum.  Its solution $\tilde{y}(t)$ reads
\begin{equation}
\tilde{y}(t) =  
a_c\, \sin\left(\Omega_c\, t + \phi_0\right) +
{a_u\, \over \sigma^2 -1 }\, \sin \Omega_u t, 
\label{app.A17}
\end{equation}
where the quantities $\Omega_u = 2\pi\,c/\lambda_u$ and $\Omega_c =
\left(q\,e\, U^{\prime\prime}/ d^2 \, m \gamma\right)^{1/2}$ are the
frequencies of, respectively, the undulator motion, i.e. the motion
along the centerline of the acoustically bent channel, and the
channeling motion due to the action of the interplanar potential.  The
amplitude of the channeling oscillations $a_c$ and the parameter
$\phi_0$ are defined by the initial conditions of the particle
entering the crystal.  As short-hand notation, $\sigma$ stands for the
ratio of the frequencies
\begin{equation}
\sigma = { \Omega_c \over \Omega_u } =
{\lambda_u\, \mu \over \pi\, d }
\label{app.A18} 
\end{equation}
with
\begin{equation}
\mu^2 =  {2 q e\,  U_o \over \varepsilon} \ll 1.
\label{app.A18a} 
\end{equation}

We remind that the dependence $\tilde{y}(t)$ defines the deviation of
the trajectory from the channel's centerline.  The total $y(t)$
dependence, according to (\ref{app.A6}), is obtained by combining
(\ref{app.A17}) with the term $a_u\, \sin (k_u c t)$.

Substituting (\ref{app.A17}) into (\ref{app.A15}) and integrating the
resulting equation one obtains the $z(t)$ dependence
\begin{eqnarray}
z(t) &=& 
c t\, \left[ 1 - {1 \over 2 \gamma^2} - 
{\xi_u^2 \over 4}\, {\sigma^4 \over (\sigma^2 -1)^2 } 
- {\mu^2 a_c^2 \over d^2} \right]
\nonumber \\
 & &- {\pi \over 4}\,{\sigma\, a_c^2 \over \lambda_u} 
\,\sin (2\Omega_c t + 2\phi_0 ) 
-{\xi_u^2 \lambda_u \over 16 \pi}\, {\sigma^4 \over (\sigma^2 -1)^2 }
\,\sin 2\Omega_u t
\label{app.A20} \\
\nonumber \\
& &-{\xi_u \, a_c \over 2}\, {\sigma^3 \over \sigma^2 -1 } 
\left[
{\cos \left((\Omega_c + \Omega_u)t + \phi_0\right) \over \sigma +1}
+
{\cos \left((\Omega_c - \Omega_u)t + \phi_0\right) \over \sigma -1}
\right].
\nonumber 
\end{eqnarray}

The analytic solution (\ref{app.A17}) allows to establish several
quantitative statements about the conditions which must be satisfied
to consider the stable channeling motion, i.e. $\tilde{y}(t) \in
[-0.5d,0.5d]$, as well as to make estimates of the relative magnitudes
of the frequencies $\Omega_u$ and $\Omega_c$.

The following conditions must be fulfilled.

\begin{itemize}
\item[(1)] $|a_c| < d/2$. 
This condition means that the amplitude of the channeling oscillations
due to the action of the interplanar potential must not exceed the
half-width of the channel.

It is equivalent to the Lindhard's condition which established the
maximal entrance angle of the particle with respect to the midplane in
a linear channel, $\Theta < \Theta_{L} = \left(2 q e U_o
/\varepsilon\right)^{1/2}$, where $\Theta_{L}$ is the Lindhard's
critical angle\citeup{Gemmell}.

This relation reflects the fact that in a linear channel the energy
$\varepsilon_y$ associated with the transverse motion is less than the
potential barrier $U_o$.  It is easy to verify that in the case of a
harmonic potential this condition is equivalent to $|a_c| < d/2$.

\item[(2)] {\it The condition for the channeling in a bent channel.}
Driven oscillations (the second term on the right-hand side of
(\ref{app.A17})) must not result in the particle's leaving the
channel.  Hence, the relation $a_u /\left( \sigma^2 -1 \right) < d/2$
must be fulfilled.  This inequality can be written in the form which
clearly exhibits the physical condition for the possibility of the
channeling process in an acoustically bent crystal.  Recalling the
definition (\ref{app.A18a}) one gets
\begin{equation}
{\varepsilon \over q \,e\,  U_{\rm max}^{\prime} \, R_{\rm min}} 
< {1 \over 1+ d/2a_u} < 1.
\label{app.A19} 
\end{equation}
The left-hand side of this relation is the ratio of the maximum
centrifugal force $\varepsilon/R_{\rm min}=\varepsilon\, k_u^2\, a_u$
and the maximum interplanar force $q \,e\, U_{\rm max}^{\prime}=4 q
\,e\, U_o/d$.  Channeling in an acoustically bent crystal can occur
only if this ration is less than 1.  For a channel bent with a
constant radius such a condition was formulated by
Tsyganov\citeup{Tsyganov}.

\item[(3)]
{\it The relationship between  the undulator 
and the channeling frequencies.}
From (\ref{app.A18}) combined with (\ref{app.A19}) one obtains
the ratio of the frequencies $\Omega_c$ and $\Omega_u$
\begin{equation}
 \sigma^2 
 > 1 + {2a_u \over d} > 1,
\label{app.A21} 
\end{equation}
which demonstrates that the frequency of the channeling motion is
always larger than that of the undulator motion.

\end{itemize}

\nonumsection{References}

\end{document}

\end{document}